\documentclass[twoside,12pt]{article}
\usepackage{epsfig}

\def\Journal#1#2#3#4{{#1} {#2} (#4) #3 }

\def\NPA{{\em Nucl. Phys.} A}

\def\PLB{{\em Phys. Lett.} B}

\def\PRL{\em Phys. Rev. Lett.}

\def\PREP{\em Phys. Rep.}

\def\PRD{{\em Phys. Rev.} D}
\def\PRC{{\em Phys. Rev.} C}

\newcommand{\bea}{\begin{eqnarray}}
\newcommand{\eea}{\end{eqnarray}}

\newcommand{\be}{\begin{equation}}
\newcommand{\ee}{\end{equation}}
\newcommand{\ba}{\begin{eqnarray}}
\newcommand{\ea}{\end{eqnarray}}
\newcommand{\del}{\partial}
\newcommand{\rth}{\frac{1}{\sqrt{3}}}
\newcommand{\rsix}{\frac{1}{\sqrt{6}}}

\newcommand{\pr}{^\prime}

\newcommand{\Gm}{\Gamma}

\newcommand{\MeV}{\textrm{ MeV}}

\newcommand{\Imass}{M_{N}}
\newcommand{\Fmass}{M_{N^{*}}}

\newcommand{\pini}{p_{i}}
\newcommand{\pfin}{P}

\newcommand{\Amp}{{\cal M}}

\begin{document}

\title{ \vspace{1cm} Photo- and Electron-Production of Mesons on Nucleons and Nuclei
}
\author{E.\ Oset$^{1}$, M.\ D\"oring$^1$, D. Strottman$^1$, D.\ Jido$^2$, 
M.\ Napsuciale$^{1}$, K.\ Sasaki$^{1}$,\\ C. A.\ Vaquera-Araujo$^{1}$, 
M.\ Kaskulov$^3$,
E.\ Hernandez$^{4}$, H.\ Nagahiro$^{5}$, S.\ Hirenzaki$^6$ \\
\\
$^1$Dep. de Fisica Teorica and IFIC, Centro Mixto Universidad de Valencia\\
 CSIC, Institutos de
Investigaci\'on de Paterna, Aptd. 22085, 46071 Valencia, Spain\\
$^2$Yukawa Institute for Theoretical Physics, Kyoto University, Kyoto, 606-8502,
Japan\\ 
$^3$Institut f\"ur Theoretische Physik, Universit\"at Giessen, Germany\\
$^4$Facultad de Ciencias, 
Universidad de Salamanca, 
E-37008 Salamanca, Spain\\
$^5$ Research Center for Nuclear Physics, Osaka University\\
 Ibaraki, Osaka 567-0047 Japan\\
$^6$Department of Physics, Nara Women's University, Nara 630-8506
Japan}
\maketitle
\begin{abstract} 
In these lectures I will show some results obtained with the chiral unitary
approach 
  applied to the photo and electroproduction of mesons. The results for
  photoproduction of $\eta \pi^0 p$ and $K^0 \pi^0 \Sigma^+$, together with
  related reactions will be shown, having with common denominator the excitation
of the $\Delta(1700)$ resonance which is one of those dynamically generated in
the chiral unitary approach. Then I will show results obtained for the 
$e^+ e^- \to \phi f_0(980)$ reaction which reproduce the bulk of the data except
for a pronounced peak, giving support to a new mesonic resonance, X(2175).
Results will also be shown for the electromagnetic form factors of the
$N^*(1535)$ resonance, also dynamically generated in this approach.
 Finally, I will show some results on the photoproduction
of the $\omega$ in nuclei, showing that present experimental results claiming a
shift of the $\omega$ mass in the medium are tied to a particular choice of
background and are not conclusive. On the other hand, the same experimental
results show unambiguously a huge increase of the $\omega$ width in the nuclear
medium. 
\end{abstract}

\section{Introduction}
Nowadays it is commonly accepted that QCD is the theory of the strong
interactions, with the quarks as building blocks for baryons and mesons, and
the gluons as the mediators of the interaction. However, at low energies typical
of the nuclear phenomena, perturbative calculations with the QCD Lagrangian are
not possible and one has to resort to other techniques to use the information of
the QCD Lagrangian. One of the most fruitful approaches has been the use of
chiral perturbation theory, $\chi PT$ \cite{xpt,ulf,ecker}. The theory introduces effective
Lagrangians which involve only observable particles, mesons and baryons,
respects the basic symmetries of the original QCD Lagrangian, particularly
chiral symmetry, and organizes these effective Lagrangians according to the
number of derivatives of the meson and baryon fields.  

The introduction of unitarity constraints in coupled channels in chiral
perturbation theory has led to unitary extensions of the theory that starting
from the same effective Lagrangians allow one to make predictions at much higher
energies . One of the interesting
consequences of these extensions is that they generate dynamically low lying
resonances, both in the mesonic and baryonic sectors. By this we mean that they
are generated by the multiple scattering of the meson or baryon components,
much  the same as the deuteron is generated by the interaction of the nucleons
through the action of a  potential, and they are not preexistent states that
remain in the large $N_c$ limit where the multiple scattering is suppressed.

\section{Baryon meson interaction}\label{baryon}

   The interaction of the octet of pseudoscalar mesons with the octet of
stable baryons has been the object of much theoretical study 
\cite{Veit:1984jr,Kaiser:1995eg,kaon,Oller:2000fj,Kaiser:1995cy,Nieves:2001wt,Inoue:2001ip,
review,GarciaRecio:2003ks,GarciaRecio:2005hy,Hyodo:2002pk,Hyodo:2006yk,bennhold,
Jido:2003cb}. One of the common features of these studies is the dynamical
generation of some resonances which appear as a consequence of the interaction of the
mesons and baryons in coupled channels, forming some kind of molecular state,
quite different for the ordinary three quark states associated to the baryons.
 The $\Lambda(1405), N*(1535),\Lambda(1670)$, etc, are examples of these states.
 I will not expose here the formalism which can be found in all these
 references ( a lecture type approach to the problem can bee seen in
 \cite{pramana}). Instead, I will briefly expose the formalism for the
 interaction of the octet of pseudoscalar mesons with the decuplet of baryons,
 which has received far less attention \cite{lutz,sarkar,GarciaRecio:2005hy}. In
 this approach one of the resonances that appears dynamically generated is the
 $\Delta(1700)$ which will play an important role in some of the reactions which
 I report here. 

    The lowest order term of the chiral Lagrangian relevant for the interaction of the baryon 
decuplet with the octet of pseudoscalar mesons is given 
by~\cite{Jenkins:1991es}\footnote{We use the metric
$g_{\mu\nu}=diag(1,-1,-1,-1)$.} 
\be
{\cal L}=-i\bar T^\mu {\cal D}\!\!\!\!/ T_\mu 
\label{lag1} 
\ee
where $T^\mu_{abc}$ is the spin decuplet field and $D^{\nu}$ the covariant derivative
given by
\be
{\cal D}^\nu T^\mu_{abc}=\del^\nu T^\mu_{abc}+(\Gm^\nu)^d_aT^\mu_{dbc}
+(\Gm^\nu)^d_bT^\mu_{adc}+(\Gm^\nu)^d_cT^\mu_{abd}
\ee
where $\mu$ is the Lorentz index, $a,b,c$ are the $SU(3)$ indices.
The vector current $\Gm^\nu$ is given by
\be
\Gm^\nu=\frac{1}{2}(\xi\del^\nu \xi^\dagger+\xi^\dagger\del^\nu \xi)
\ee
with
\be
\xi^2=U=e^{i\sqrt{2}\Phi/f}
\ee 
where $\Phi$ is the ordinary 3$\times$3 matrix of fields for the pseudoscalar 
mesons~\cite{xpt} and $f=93$ MeV. We deal here only about the $S$-wave 
part of the baryon meson interaction and  take for the Rarita-Schwinger fields $T_\mu$ the
representation $Tu_\mu$ from ref.~\cite{bookericson,holstein} with $u_\mu$ the
Rarita-Schwinger spinor.

Let us recall the identification of the $SU(3)$ component
of $T$ to the physical states~\cite{savage,lutz3}:

$T^{111}=\Delta^{++}$, $T^{112}=\rth\Delta^{+}$, $T^{122}=\rth\Delta^{0}$,
$T^{222}=\Delta^{-}$, $T^{113}=\rth\Sigma^{*+}$, $T^{123}=\rsix\Sigma^{*0}$,
$T^{223}=\rth\Sigma^{*-}$,  $T^{133}=\rth\Xi^{*0}$,
$T^{233}=\rth\Xi^{*-}$, $T^{333}=\Omega^{-}$.

For a meson of incoming (outgoing) momenta $k(k\pr)$ one obtains  the 
simple form for the $S$-wave transition amplitudes, similar to~\cite{kaon},
\be
V_{ij}=-\frac{1}{4f^2}C_{ij}(k^0+k^{\pr 0}).
\label{poten}
\ee 

The coefficients $C_{ij}$ for reactions with all possible values of
strangeness $(S)$ and charge $(Q)$ are given in \cite{sarkar}. This interaction
plays the role of the kernel in the Bethe Salpeter equation which is used to 
construct the scattering t-matrix between the different channels. In the next
section we approach this point.

 \section{Unitarized chiral perturbation theory: N/D or dispersion relation
method}

  One can find a systematic and easily comprehensible derivation 
 of the  ideas of the N/D method applied for the first time to the meson baryon system in
 \cite{Oller:2000fj} in the context of chiral dynamics, which we reproduce 
 here below and which follows closely
 the similar developments used before in the meson meson interaction \cite{nsd}.
 One defines the transition $T-$matrix as $T_{i,j}$ between the coupled channels which couple to
 certain quantum numbers. For instance in the case of  $\bar{K} N$ scattering studied in
 \cite{Oller:2000fj} the channels with zero charge are $K^- p$, $\bar{K^0} n$, $\pi^0 \Sigma^0$,$\pi^+
 \Sigma^-$, $\pi^- \Sigma^+$, $\pi^0 \Lambda$, $\eta \Lambda$, $\eta \Sigma^0$, 
 $K^+ \Xi^-$, $K^0 \Xi^0$.
 Unitarity in coupled channels is written as
 
\begin{equation} 
Im T_{i,j} = T_{i,l} \rho_l T^*_{l,j}
\end{equation}
where $\rho_i \equiv 2M_l q_i/(8\pi W)$, with $q_i$  the modulus of the c.m. 
three--momentum, and the subscripts $i$ and $j$ refer to the physical channels. 
 This equation is most efficiently written in terms of the inverse amplitude as
\begin{equation}
\label{uni}
\hbox{Im}~T^{-1}(W)_{ij}=-\rho(W)_i \delta_{ij}~,
\end{equation}
The unitarity relation in Eq. (\ref{uni}) gives rise to a cut in the
$T$--matrix of partial wave amplitudes, which is usually called the unitarity or right--hand 
cut. Hence one can write down a dispersion relation for $T^{-1}(W)$ 
\begin{equation}
\label{dis}
T^{-1}(W)_{ij}=-\delta_{ij}\left\{\widetilde{a}_i(s_0)+ 
\frac{s-s_0}{\pi}\int_{s_{i}}^\infty ds' 
\frac{\rho(s')_i}{(s'-s)(s'-s_0)}\right\}+{\mathcal{T}}^{-1}(W)_{ij} ~,
\end{equation}
where $s_i$ is the value of the $s$ variable at the threshold of channel $i$ and 
${\mathcal{T}}^{-1}(W)_{ij}$ indicates other contributions coming from local and 
pole terms, as well as crossed channel dynamics but {\it without} 
right--hand cut. These extra terms
are taken directly from $\chi PT$ 
after requiring the {\em matching} of the general result to the $\chi PT$ expressions. 
Notice also that 
\begin{equation}
\label{g}
g(s)_i=\widetilde{a}_i(s_0)+ \frac{s-s_0}{\pi}\int_{s_{i}}^\infty ds' 
\frac{\rho(s')_i}{(s'-s)(s'-s_0)}
\end{equation}
is the familiar scalar loop integral.

One can further simplify the notation by employing a matrix formalism. 
Introducing the 
matrices $g(s)={\rm diag}~(g(s)_i)$, $T$ and ${\mathcal{T}}$, the latter defined in 
terms 
of the matrix elements $T_{ij}$ and ${\mathcal{T}}_{ij}$, the $T$-matrix can be written as:
\begin{equation}
\label{t}
T(W)=\left[I-{\mathcal{T}}(W)\cdot g(s) \right]^{-1}\cdot {\mathcal{T}}(W)~.
\end{equation}
which can be recast in a more familiar form as 
 \begin{equation}
\label{ta}
T(W)={\mathcal{T}}(W)+{\mathcal{T}}(W) g(s) T(W)
\end{equation}
Now imagine one is taking the lowest order chiral amplitude for the kernel
${\mathcal{T}}$ as done in
\cite{Oller:2000fj}. Then the former equation is nothing but the Bethe Salpeter equation with the
kernel taken from the lowest order Lagrangian and  factorized  on  shell, the same
approach followed in \cite{kaon}, where different arguments were used to justify the on shell
factorization of the kernel.

The on shell factorization of the kernel, justified here with the N/D method,
renders the set of coupled Bethe Salpeter integral equations a simple set of
algebraic equations.

In the case of the interaction of the octet of pseudoscalar mesons with the
baryon decuplet, one of the resonances which appears dynamically generated 
 is the $\Delta(1700)$, which is 
built from the
coupled channels $\pi \Delta$, $\eta \Delta$  and $K \Sigma(1535)$. 
The study of the amplitudes
around the pole allows one to get the mass, the width, and through the residues
at the pole, the coupling of the resonance to the different channels. Another
one of such states is the $\Lambda(1520)$ which couples to the 
$\pi \Sigma(1385)$ and $K \Xi(1530)$ channels, although it has also an
appreciable coupling to the $\pi \Sigma$ and $\bar{K} N$ channels, 
responsible for its width, and to which it couples in D-waves \cite{Roca:2006sz}.

\section{Clues to the nature of the $\Delta(1700)$ resonance from pion and
photon induced reactions}

\begin{figure}[htb]
\begin{center}
\includegraphics[width=4.5cm]{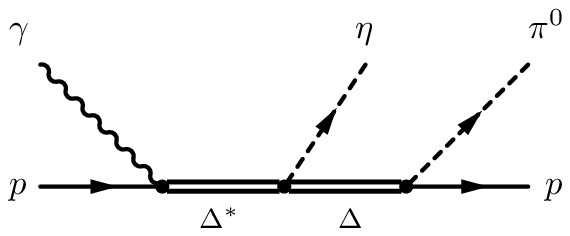}
\hspace{1cm}
\includegraphics[width=4.5cm]{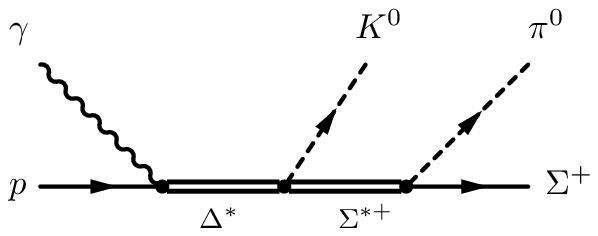}
\caption{Tree level processes from the decay of the $\Delta^*(1700)$ 
to $\eta\Delta(1232)$ and $K^0\Sigma^{*+}$.}
\label{fig:background}
\end{center}
\end{figure}

  As mentioned in the previous section, the $\Delta(1700)$ is a dynamically
generated resonance from the   $\pi \Sigma$ and $\bar{K} N$
channels with the interaction discussed in section 2. The couplings to these
channels are evaluated in \cite{sarkar}. If we study the 
$\gamma p \to \pi^0 \eta p$ or the  $\gamma p \to \pi^0 K^0 \Sigma^+$ reactions
it was found in \cite{etapi} and \cite{kpi} that the  mechanisms of
$\Delta(1700)$ excitation with posterior decay of the $\Delta(1700)$ into 
$\eta \Delta$ and $K^0 \Sigma(1385)$ were dominant in these reactions at low
energies. The corresponding mechanisms are depicted in fig \ref{fig:background}.

\begin{figure}[htb]
\begin{center}
\includegraphics[width=8cm]{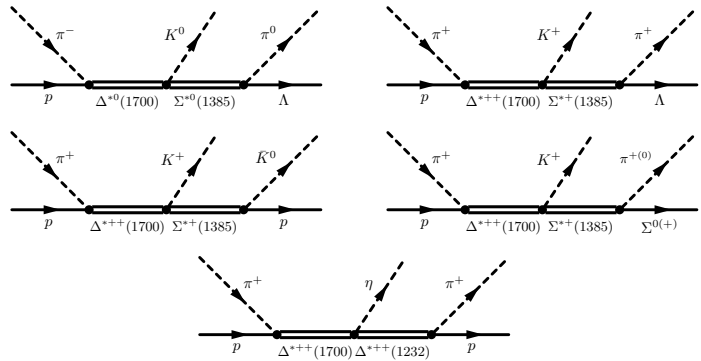}
\caption{Tree level contributions for the pion-induced strangeness production. }
\label{fig:tree_level_new}
\end{center}
\end{figure}

By analogy, reactions with the same final state induced by other probes should
show similar features. In fig. \ref{fig:tree_level_new} we show the corresponding Feynman
diagrams for pion induced reactions.

\begin{figure}[htb]
\begin{center}
\includegraphics[width=10cm]{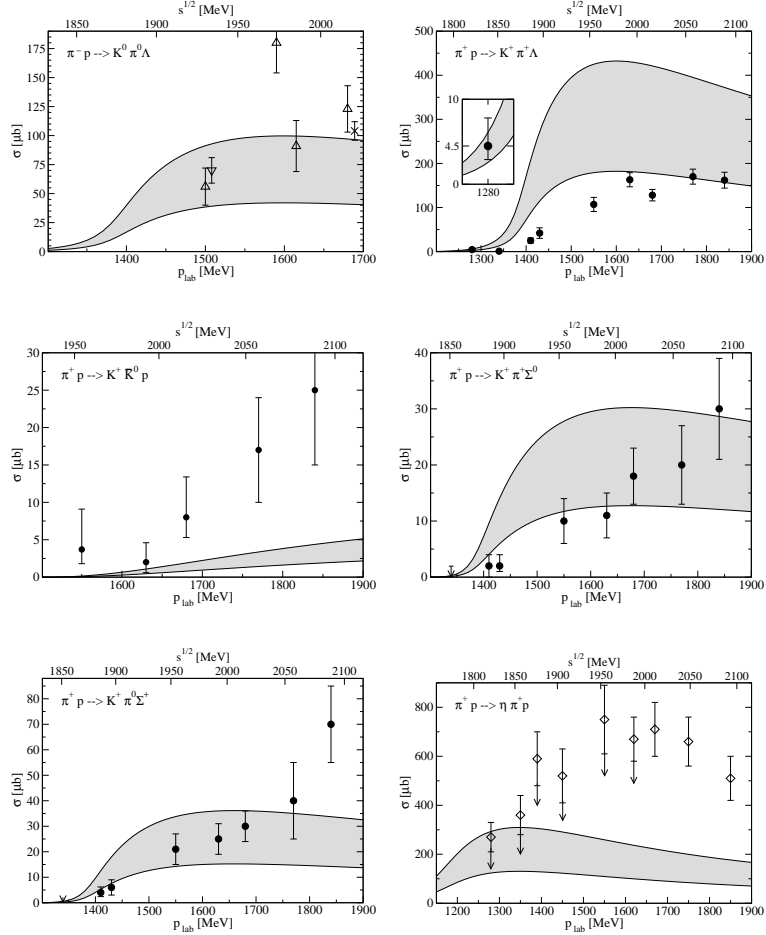}
\caption{Total cross sections for the pion-induced reactions. Data are from 
\cite{dahl} (triangles up),
\cite{curtis} (triangle down),
\cite{Thomas:1973uh} (cross),
\cite{Hanson:1972zz} (dots), 
\cite{Grether:1973sz} (diamonds).
For the latter data, it is indicated that these are upper limits 
below $p_{\rm}=1.67$ GeV as in Ref. \cite{Grether:1973sz}.}
\label{fig:sigma}
\end{center}
\end{figure}

  In fig. \ref{fig:sigma} we show the results for the cross sections for several
  pion induced reactions which were evaluated in \cite{kpi}. The overall
  agreement with the data is fair. However, this fair agreement gains
 more strength after the following considerations.
   The $\Delta(1700)$ is assumed to be a member of a decuplet in the PDG. If one
assumes this, it is easy, using SU(3) Clebsch Gordan coefficients, to see that the
 couplings to the 
$\Delta\pi$, $\Sigma^* K$, $\Delta\eta$ states in $I=3/2$ are proportional 
to $\sqrt{5/8}$, $\sqrt{1/4}$, $\sqrt{1/8}$, respectively.  The squares of 
these coefficients are proportional to $1$, $2/5$, $1/5$, respectively, compared
 to the squares of the coefficients of the dynamically generated resonance,
  $1$, $11.56$, $4.84$.  With respect 
   to the decuplet assumption of the PDG one obtains factors $27.5$ and $24$ 
   larger for the square of the couplings to $\Sigma^*K$ and $\Delta\eta$, 
   respectively.   In some cross sections involving the square of
these numbers the differences would be huge. With this perspective, the fair
agreement found with experiment has a more significant value.

\begin{figure}[htb]
 \begin{center}
\includegraphics[width=14cm]{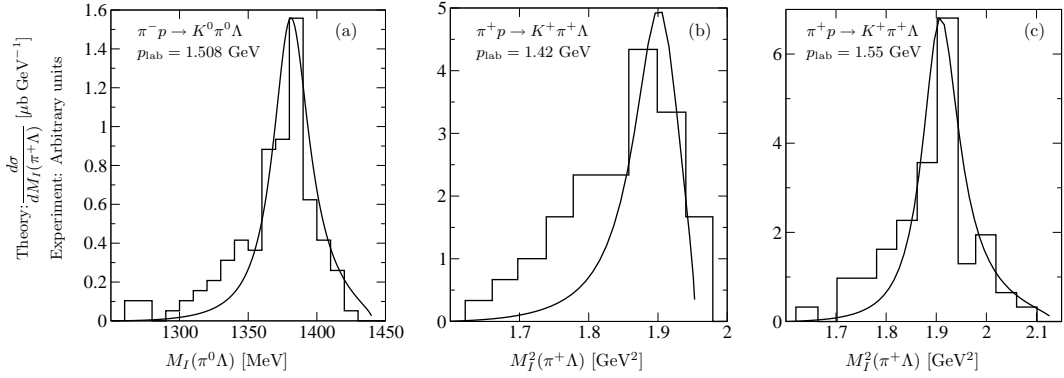}
\caption{Invariant mass spectra for the $\pi^-p\to K^0\pi^0\Lambda$ and $\pi^+ p \to K^+ \pi^+ \Lambda$ reactions. 
Experimental distributions (arbitrary units) are from \cite{curtis} and \cite{Hanson:1972zz}, respectively.}
\label{fig:dsdmi}
\end{center}
\end{figure}

As we indicated, the case for $K^0 \pi^0 \Lambda$ is dominated in our model
by $K^0\Sigma(1385)$ production. This is the case experimentally, as one can see
in fig. \ref{fig:dsdmi}. 

\begin{figure}[htb]
\includegraphics[width=5cm]{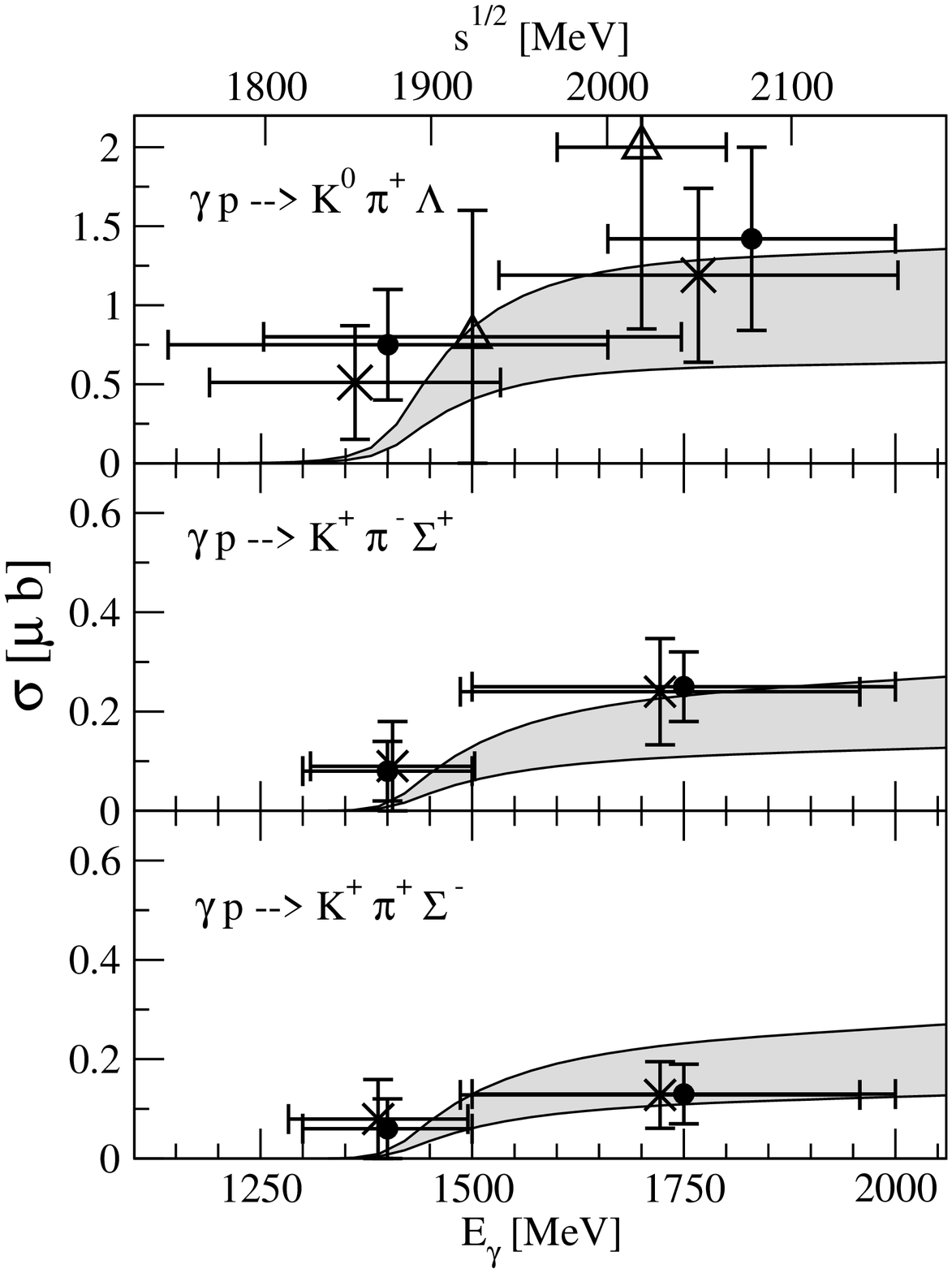}
\hspace{0.5cm}
\includegraphics[width=4.8cm]{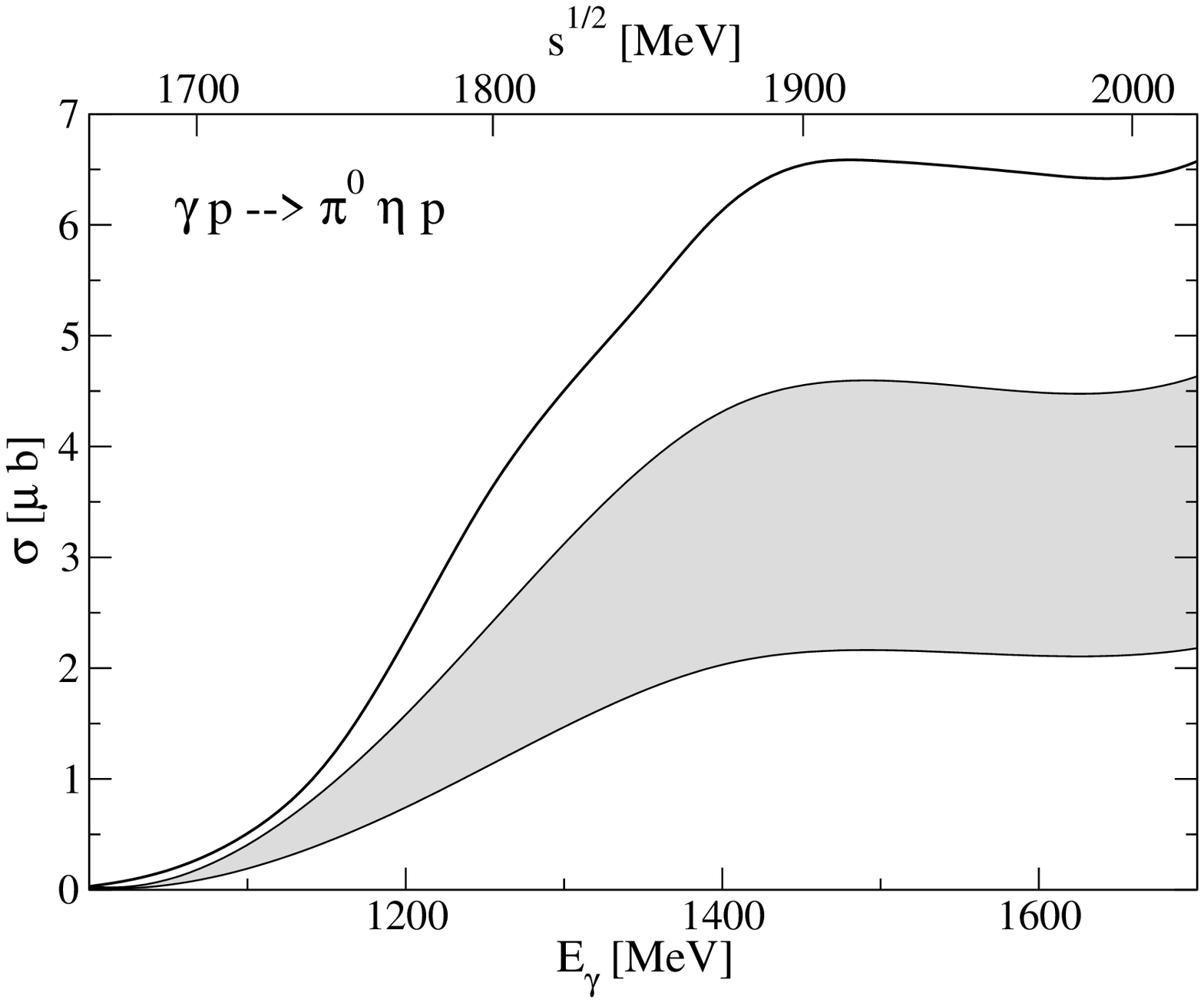}
\hspace*{0.5cm}
\includegraphics[width=5.1cm]{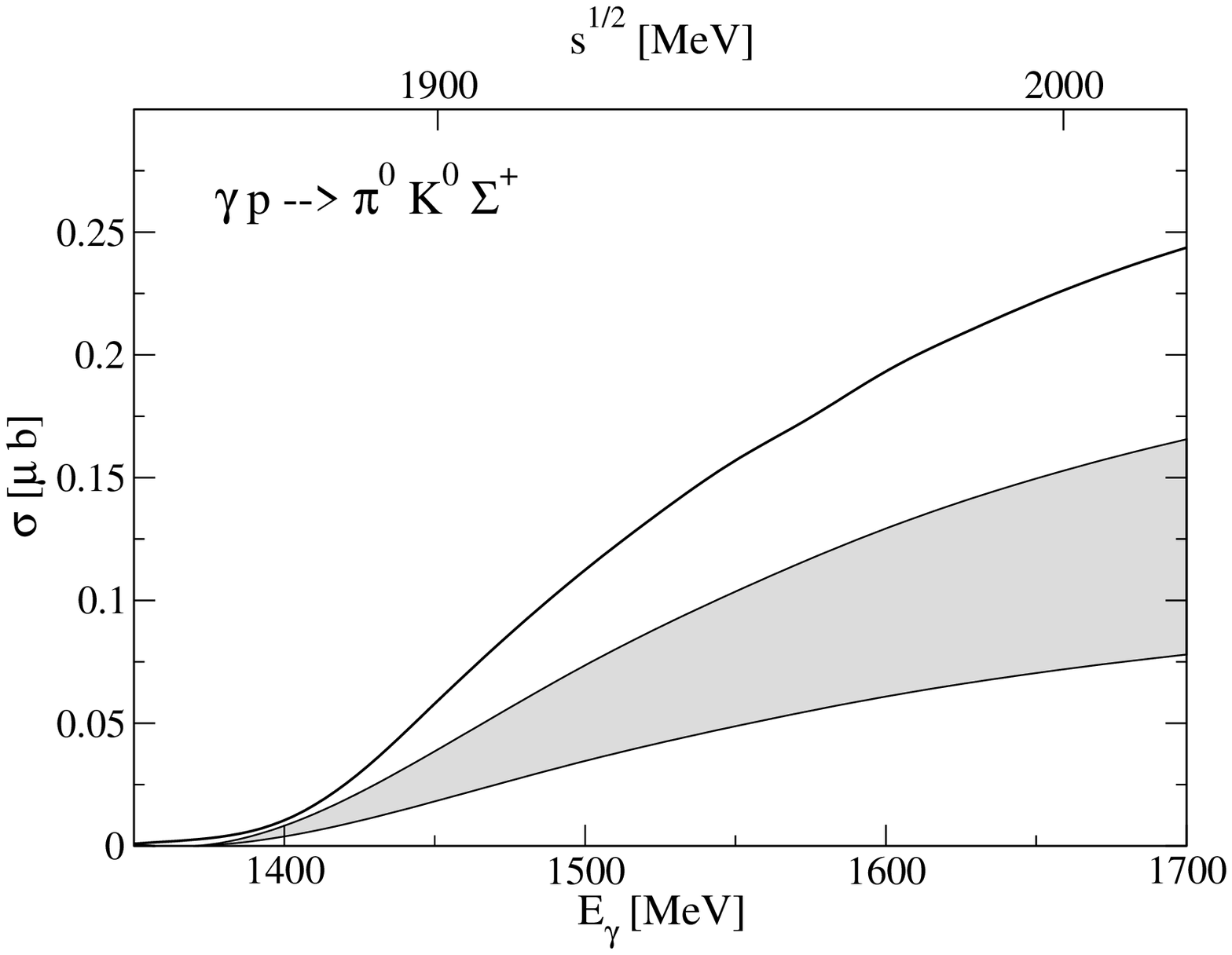}
\caption{Photoproduction of strange and $\eta$ particles. Data are from 
\cite{Erbe:1970cq} (dots), \cite{Erbe:2} (crosses), and \cite{cmc} (triangles up). The latter data are the sum of $K\Lambda\pi$ and $K\Sigma\pi$ final states. 
The two lower plots show the update of the predictions from Ref. \cite{etapi}
(gray bands) compared to the results from \cite{kpi} (solid lines).}
\label{fig:photo}
\end{figure}

   The case of photoproduction is also instructive. In fig. \ref{fig:photo} we
can see that there is a good agreement with the data, though the experimental
errors are large. The precision of the experiment has been significantly
improved in recent preliminary measurements at ELSA, and the agreement of our
results with the data is very good \cite{wieland}. Finally, in the same figure
we can see the predictions for the $\gamma p \to \pi^0 \eta p$ and 
$\gamma p \to K^0 \pi^0 \Sigma^+$ production which have also been measured
recently at ELSA \cite{mariana} and the agreement with the data is also good,
as well as with the data from \cite{nakabayashi} and recent data by the GRAAL
collaboration on $\gamma p \to \pi^0 \eta p$ \cite{graal}, where is addition we
find also good agreement with the asymmetries \cite{graal}. It is worth stating
that in the absence of the $\Delta(1700)$ intermediate state (the model of
\cite{etapi} has more terms than just the $\Delta(1700)$ excitation), the
results for the asymmetries are in sheer disagreement with the data. 

\section{The $e^+ e^- \to \phi f_0(980)$ and clues for a new mesonic resonance 
X(2175)}

\begin{figure}[htb]
\begin{center}
\includegraphics[
height=4.50in,
width=3.75in
]{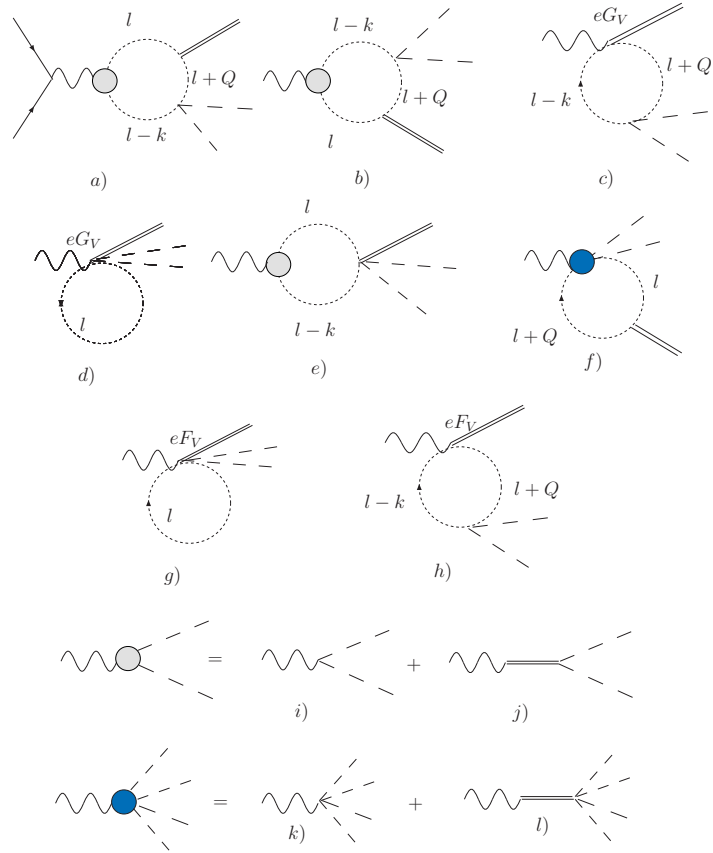}
\end{center}
\caption{Feynman diagrams for $e^{+}e^{-}\rightarrow\protect\phi\protect\pi%
\protect\pi$ in $R\protect\chi PT$.}
\label{FD}
\end{figure}

In  recent experiments \cite{BBY,BBX} the $e^+ e^- \to \phi \pi^+ \pi^-$ and 
$e^+ e^- \to \phi \pi^0 \pi^0$ cross sections were measured and a pronounced peak
was found for the invariant mass of the two pions in the region of the 
$f_0(980)$ scalar meson resonance. By looking at the energy dependence of the
 $e^+ e^- \to \phi f_0(980)$ a neat peak was found on top of a smooth 
 background which was identified in \cite{BBY,BBX} with a new resonance X(2175). 
 We have attempted to reproduce these data by recalling the model used for the 
 $\phi \to \gamma f_0(980)$ in \cite{eugenio} (see also other works in the same
 direction \cite{others}), which would be the time
 reversal reaction of the present one, with the novelty of having virtual
 photons.  We have used the Feynman diagrams of fig. \ref{FD}, which involve
  kaons as intermediate states. We observed that because one has larger energies
  than in the case of the $\phi$, we had to introduce also $K^*$ in the
  intermediate states. The production of the $f_0(980)$ is done automatically
in our approach, since this resonance is also dynamically generated by the
interaction of meson-meson in coupled channels. Hence, the t-matrices that we
implement in the loops automatically generate this resonance which does not have
to be introduced by hand. The details can be seen in \cite{mauro}.

The evaluation of the amplitude requires to work out the loop tensor $%
T_{\mu \nu }^{abc}$, such that the amplitude is 
$ T_{\mu \nu }^{abc} \epsilon_{\mu}(\gamma) \epsilon_{\nu}(\phi)$. 
It can be  shown that $T_{\mu \nu }^{abc}$ is
finite and gauge invariant. The most general form of this tensor is 
\begin{equation}
T_{\mu \nu }^{abc}=a\ g_{\mu \nu }+b\ Q_{\mu }Q_{\nu }+c\ Q_{\mu }k_{\nu
}+d\ k_{\mu }Q_{\nu }+e\ k_{\mu }k_{\nu }
\end{equation}%
where $\ a,b,c,d,e$ are form factors and k and Q are
the momenta of the photon and the $\phi$ respectively. Gauge invariance requires 
\begin{equation}
k^{\mu }T_{\mu \nu }^{abc}=\left( a+ck\cdot Q+ek^{2}\right) k_{\nu
}+(bk\cdot Q+dk^{2})Q_{\nu }=0,
\end{equation}%
imposing the following relations among the form factors%
\begin{equation}
a=-c\ k\cdot Q-e\ k^{2},\qquad b\ k\cdot Q=-d\ k^{2},
\end{equation}%
thus $T_{\mu \nu }^{abc}$ has the following explicitly gauge invariant form 
\begin{equation}
T_{\mu \nu }^{abc}=-c(Q\cdot k\ g_{\mu \nu }-Q_{\mu }k_{\nu })-\frac{d}{%
k\cdot Q}\ (k^{2}Q_{\mu }-k\cdot Qk_{\mu })Q_{\nu }-e\ (k^{2}g_{\mu \nu
}-k_{\mu }k_{\nu }).  \label{git}
\end{equation}%
The second term vanishes upon contraction with $\epsilon_{\nu}(\phi)$ and
 we are left only with two form factors 
\begin{equation}
T_{\mu \nu }^{abc}=-c(Q\cdot k\ g_{\mu \nu }-Q_{\mu }k_{\nu })-e\
(k^{2}g_{\mu \nu }-k_{\mu }k_{\nu }).  \label{tmunugen}
\end{equation}

\begin{figure}[t]
%
%
\includegraphics[height=3.50in,width=4.00in]{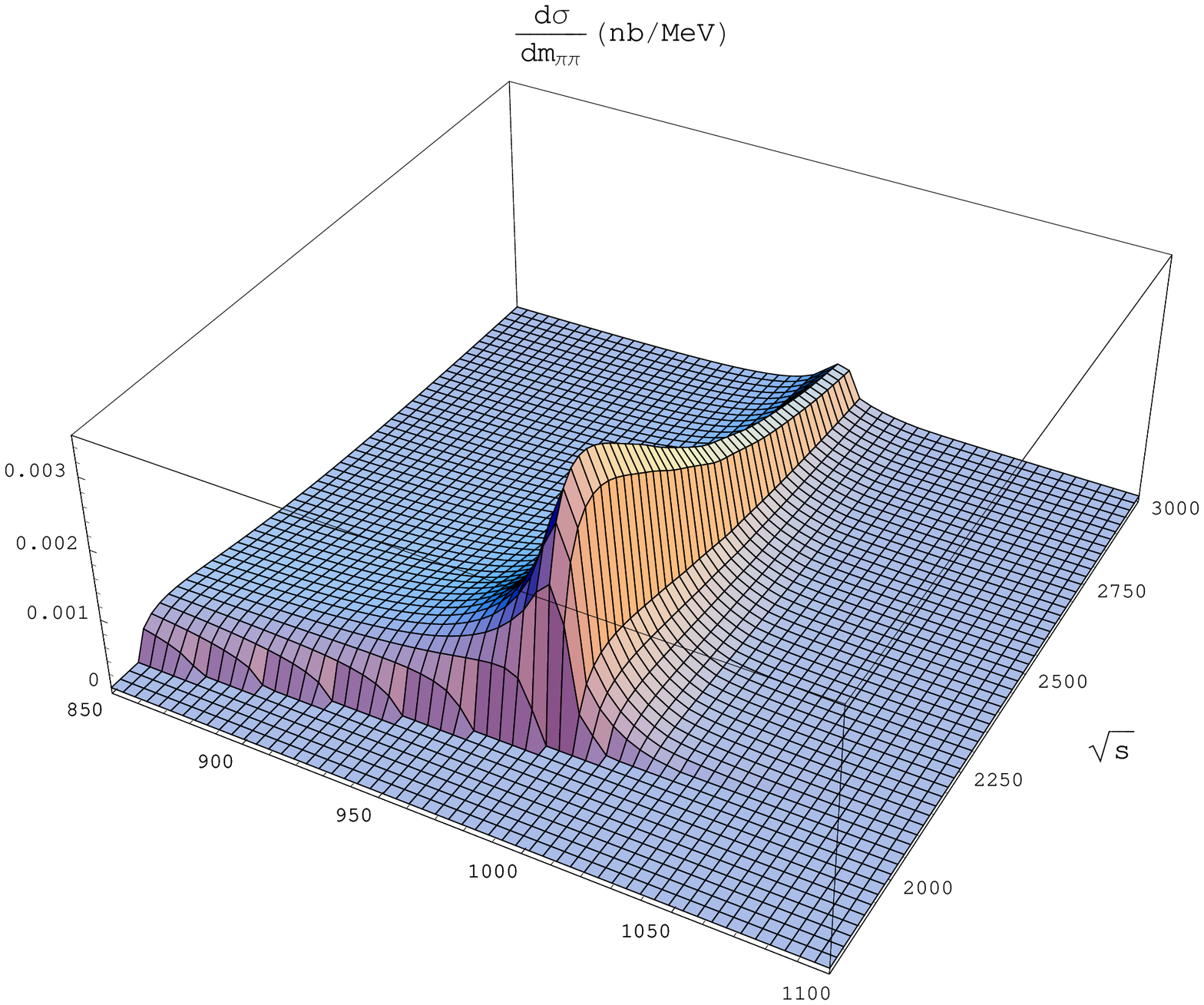}
\hspace{0.01cm}
\includegraphics[height=3.50in,width=3.00in]{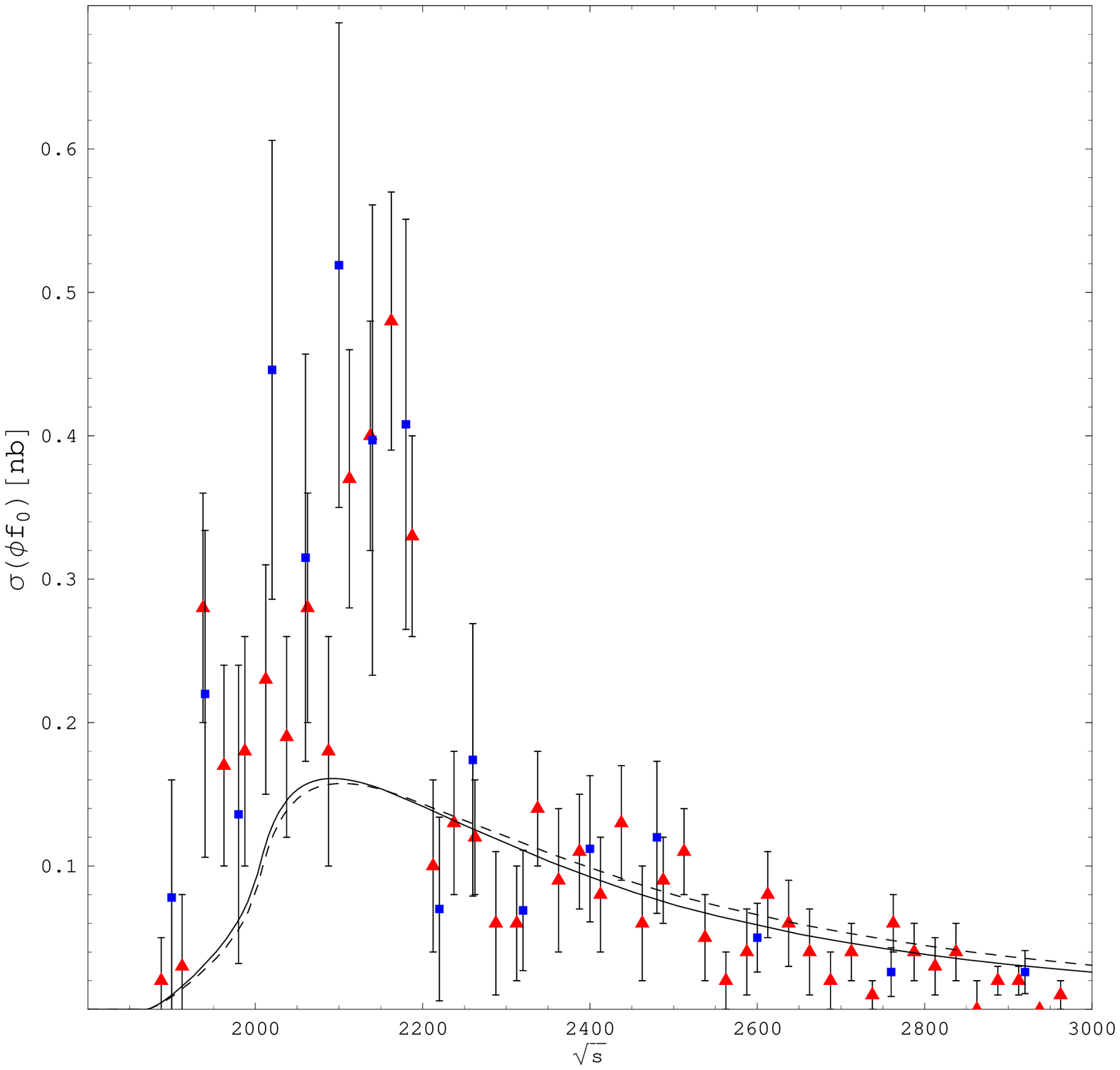}
\caption{ Left: Differential cross section as a function of the dipion invariant
mass and of the center of mass energy. Right: Cross section for $e^{+}e^{-}\rightarrow\protect\phi\left[ \protect%
\pi\protect\pi\right] _{I=0}$, integrated in the $m_{\protect\pi\protect\pi%
}=850-1100~ MeV$ range, as a function of $\protect\sqrt{s}$ including all contributions.
Experimental points from Ref. \protect\cite{BBX}, triangled (boxed) points
correspond to charged (neutral) pions.}
\label{dsig}
\end{figure}

The results for the double distribution in total energy and invariant mass of
two pions can be seen in fig. \ref{dsig} (left), where the crest due to the $f_0(980)$
excitation is clearly visible. The integrated cross section around the crest of
the $f_0(980)$, as done in the experimental analysis, is shown in  fig. \ref{dsig}
(right). We can see there that we can reproduce the bulk of the data, in
strength and shape, but the peak around 2175 MeV is not reproduced. There 
is an extra cross section with an approximate Breit Wigner shape which 
suggest the
excitation of a resonance which decays into $\phi f_0(980)$. This was the
conclusion of \cite{BBY,BBX} based on a best fit to the data. Our study offers
and extra support to this claim by showing that known dynamics of the processes
can explain the background, but not the peak.

\section{Electroproduction and form factors of the $N^*(1535)$}

\begin{figure}[t]
\hspace{1cm}
\epsfxsize= 4cm
\epsfbox{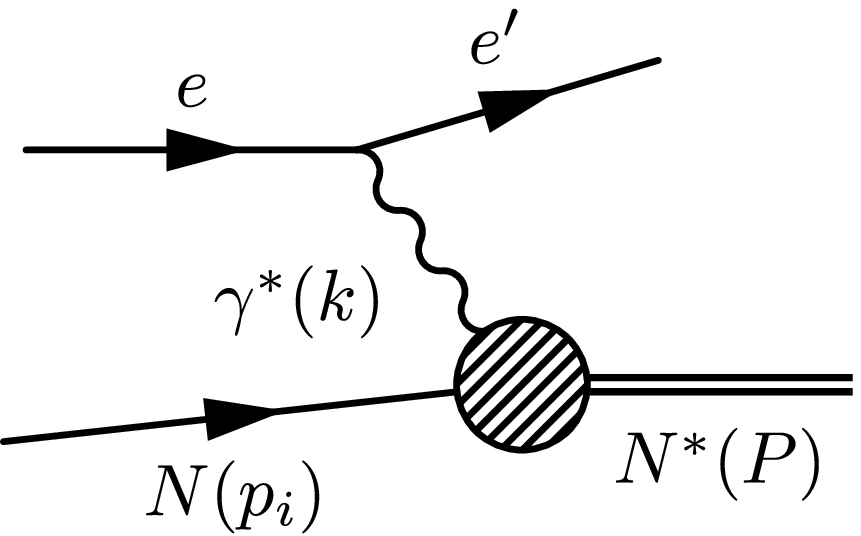}
\hspace{3cm}
\epsfxsize= 8.5cm
  \epsfbox{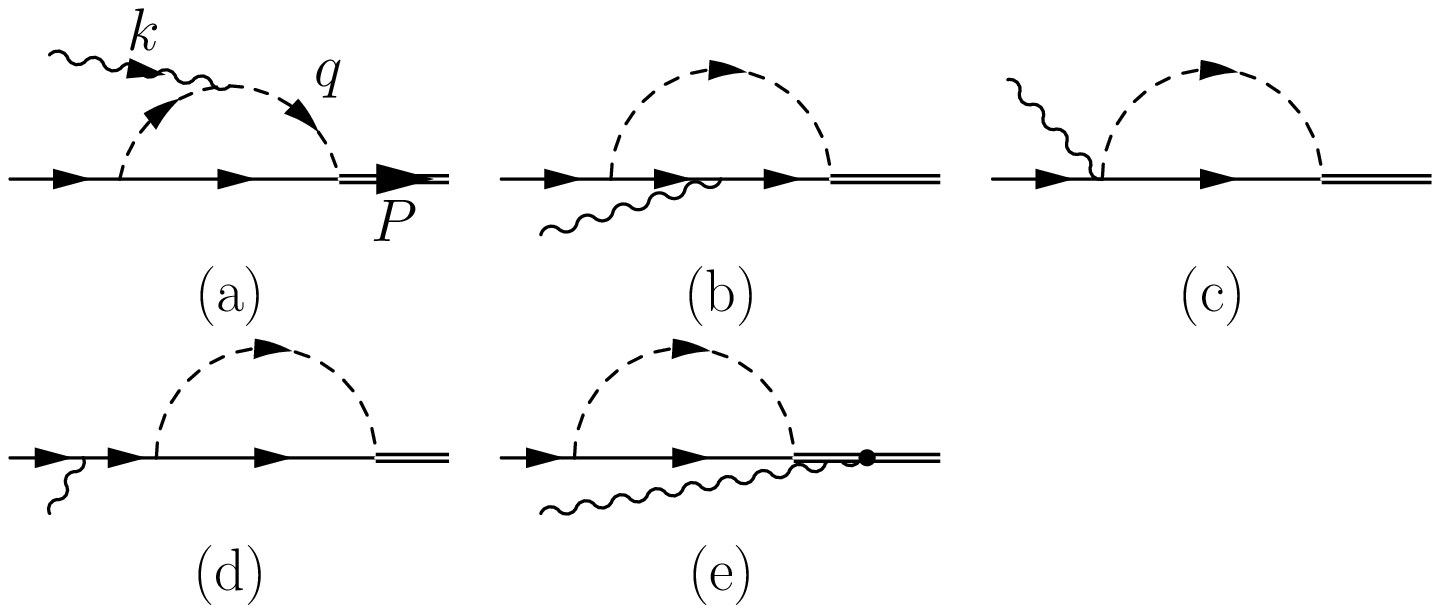}

  \caption{ Letf: Kinematics of the electroproduction of $N(1535)$.
  Right: Feynman diagrams for the transition form factor of $N(1535)$ at one 
  loop level. The solid, dashed, wavy and double lines denote octet baryons,
   mesons, photon and $N(1535)$, respectively. \label{fig:FeynmanDiagram}}
\end{figure}

 We now show some new results on the form factors of the $N^*(1535)$ obtained
 from the electroproduction of this resonance, see fig.
 \ref{fig:FeynmanDiagram}.

There are two independent amplitudes for the electro-transition 
from $J^{P}=1/2^{+}$ to $1/2^{-}$, $A_{1/2}$ and $S_{1/2}$, which are defined 
in terms of the transition electric current $J_{\mu}$ by
\begin{eqnarray}
   A_{1/2} &=& \sqrt \frac{2\pi \alpha}{q_{R}} \frac{1}{e} 
   \langle N^{*}, J_{z}=\frac{1}{2} | \epsilon^{(+)}_{\mu} J^{\mu} | N, S_{z}=- \frac{1}{2} \rangle \\
   S_{1/2} &=& \sqrt \frac{2\pi \alpha}{q_{R}} \frac{1}{e}\frac{|\vec k|}{\sqrt{Q^{2}}} 
   \langle N^{*}, J_{z}=\frac{1}{2} | \epsilon^{(0)}_{\mu} J^{\mu} | N, S_{z}= \frac{1}{2} \rangle \ \ \
\end{eqnarray}
with the fine structure constant $\alpha = e^{2}/4\pi$, the energy equivalent of a 
real photon $q_{R} =(W^{2}-\Imass^{2})/(2W)$ and the photon-nucleon 
center-of-mass energy $W\equiv \sqrt{\pfin^{2}}$. The polarization vectors of 
the photon, $\epsilon_{\mu}$, are given by
\begin{equation}
   \epsilon^{\pm}_{\mu} = \frac{1}{\sqrt 2} (0,\mp 1, -i ,0); ~~~~~~
   \epsilon^{0}_{\mu} = \frac{1}{\sqrt {Q^{2}}} (k ,0, 0, -k^{0})    
\end{equation}
with $Q^{2} = - k^{2}$, where we take the CM momenta 
$\vec k$ and $\vec \pini$ along the $z$ axis. 

It can be seen, using the equation of motion for the spinors and conservation of
momenta, that the 
transition current $J^{\mu}$ can be written, in general, by the following three 
Lorentz scalar amplitudes: 
\begin{equation}
   J^{\mu} = (\Amp_{1} \gamma^{\mu} + \Amp_{2} P^{\mu} + \Amp_{3} k^{\mu})\gamma_{5} . \label{eq:Jmu}
\end{equation}

The gauge invariance $  k \cdot J = 0$, tells us that there are only two independent amplitudes 
among these three amplitudes, $\Amp_{i}$,  giving the following relation:
\begin{equation}
   (\Fmass+\Imass) \Amp_{1} + k\cdot P \Amp_{2} + k^{2} \Amp_{3} = 0  \label{eq:iden} \ .
\end{equation}

Since the $N^*(1535)$ is also dynamically generated in our approach, we obtain
the electroproduction amplitude by coupling the photon to the meson-baryon
components that build up this resonance \cite{Inoue:2001ip}. The mechanisms are
depicted in fig.\ref{fig:FeynmanDiagram} (right), which produce a gauge invariant set of diagrams.

\begin{figure}[htb]
\epsfxsize= 8.5cm
  \epsfbox{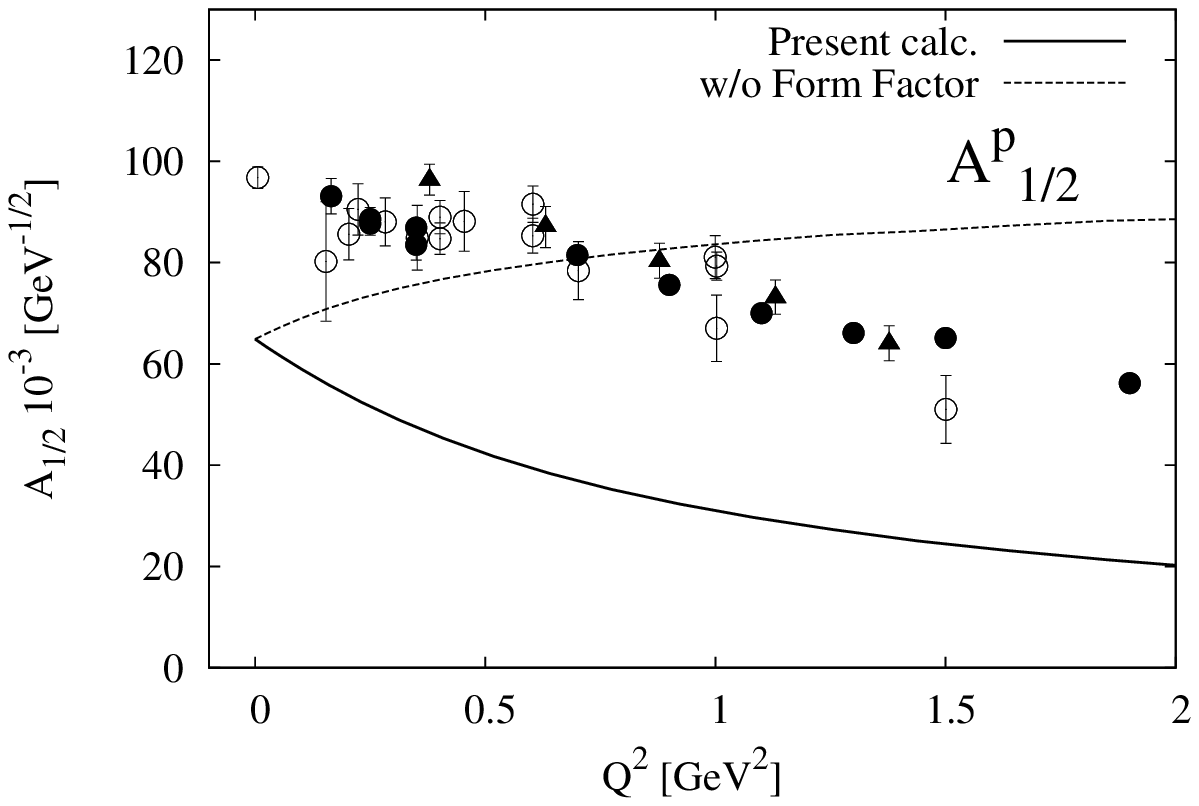}
\hspace{0.05cm}
 \epsfxsize= 8.5cm
  \epsfbox{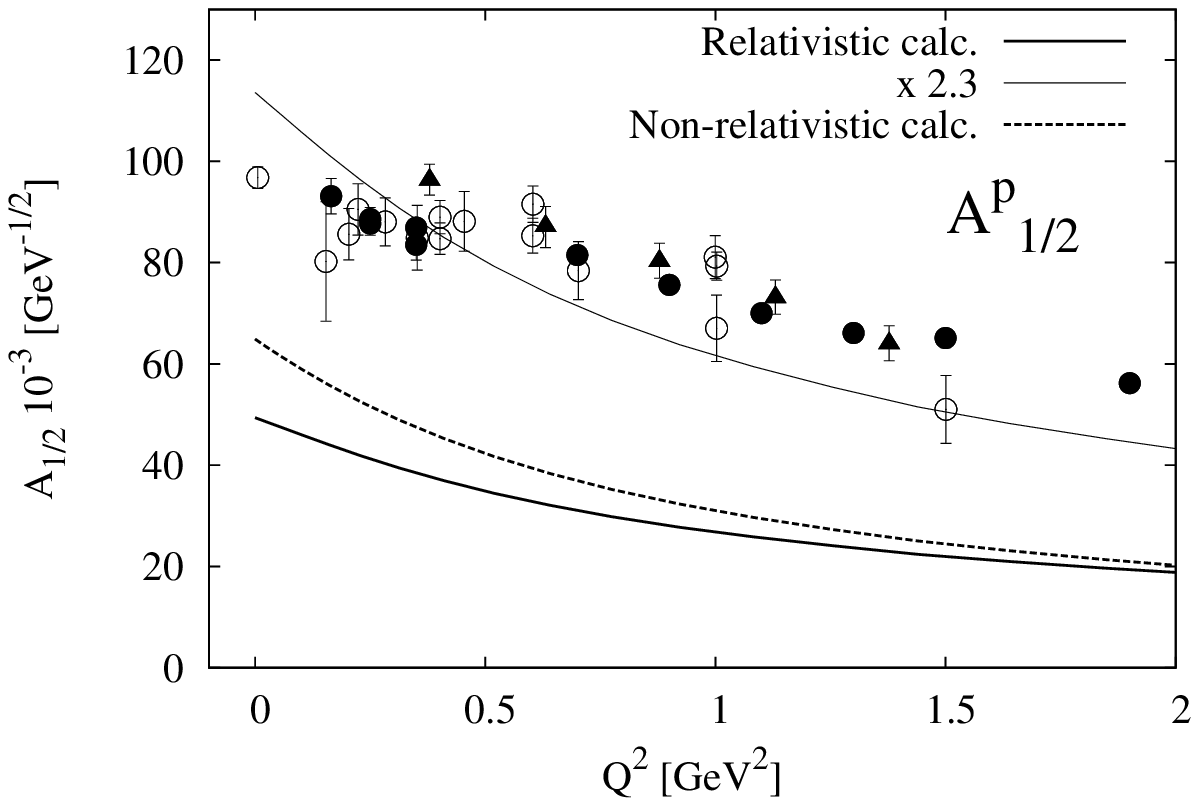} 
\caption{Left: Modulus of the $A_{1/2}$ helicity amplitude for the proton resonance 
as a function of $Q^{2}$ with $W=1535 \MeV$ calculated in the non-relativistic
 formulation. The solid (dotted) line shows the $A_{1/2}^{p}$ amplitude with 
 (without) the form factor of the meson inside the loops.  
Marks with error bars are experimental data normalized by the $N^{*}$ full 
width $\Gamma_{N^{*}}=150$ MeV and the $N^{*} \rightarrow \eta N$ branching 
ratio $b_{\eta}=0.55$. Filled triangles and circles are results of the CLAS 
collaboration taken from Refs.\cite{Thompson:2000by} and  \cite{Denizli:2007tq},
 respectively.  Open circles show results of 
 Refs.\cite{Krusche:1995nv,Brasse:1977as,Beck:1974wd,Breuker:1978qr}. 
  Righ: Modulus of the 
 $A_{1/2}$ helicity amplitude for the proton resonance as a function of $Q^{2}$ with $W=1535 \MeV$ in relativistic formulation. 
The solid (dotted) line shows the $A_{1/2}^{p}$ amplitude with (without) the form factor of the meson inside the loops.  
\label{fig:relaA1/2}
%
\label{fig:A1/2proton}}
\end{figure}

\begin{figure}[htb]
\epsfxsize= 8.5cm
  \epsfbox{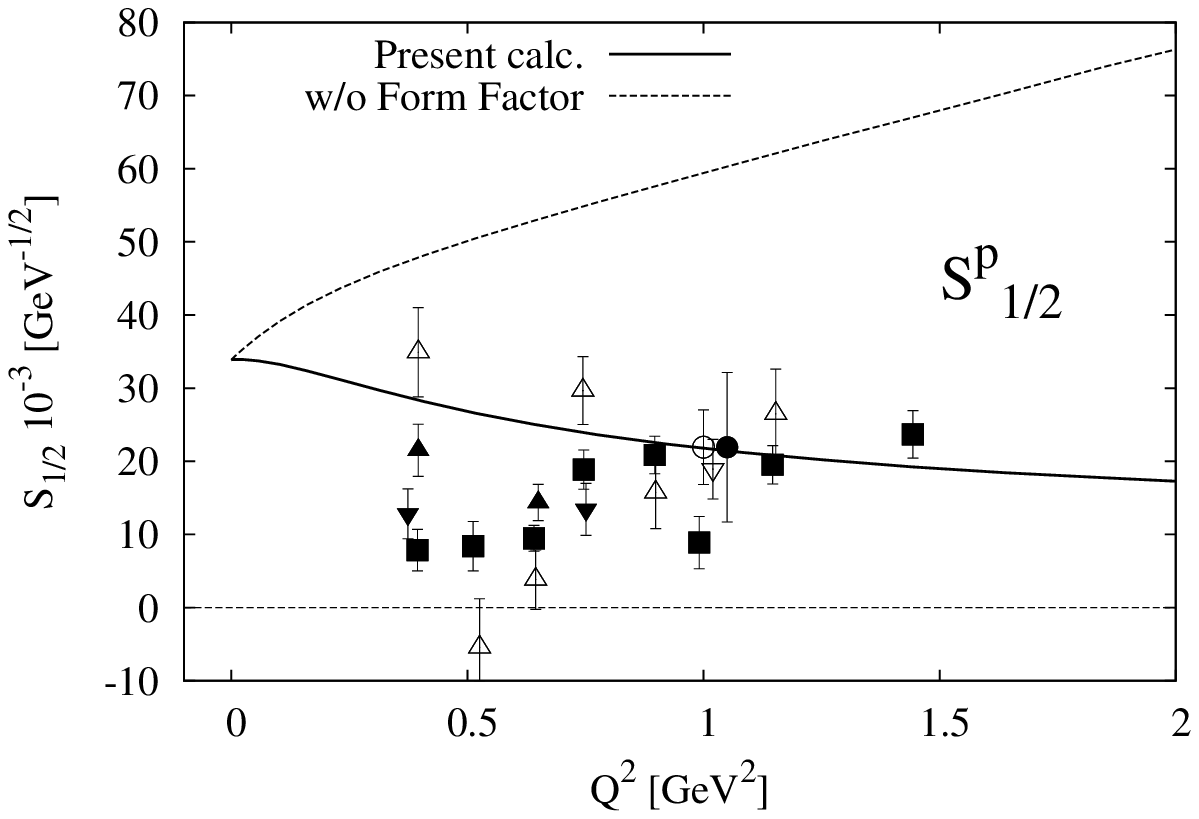}
\hspace{0.15cm}
 \epsfbox{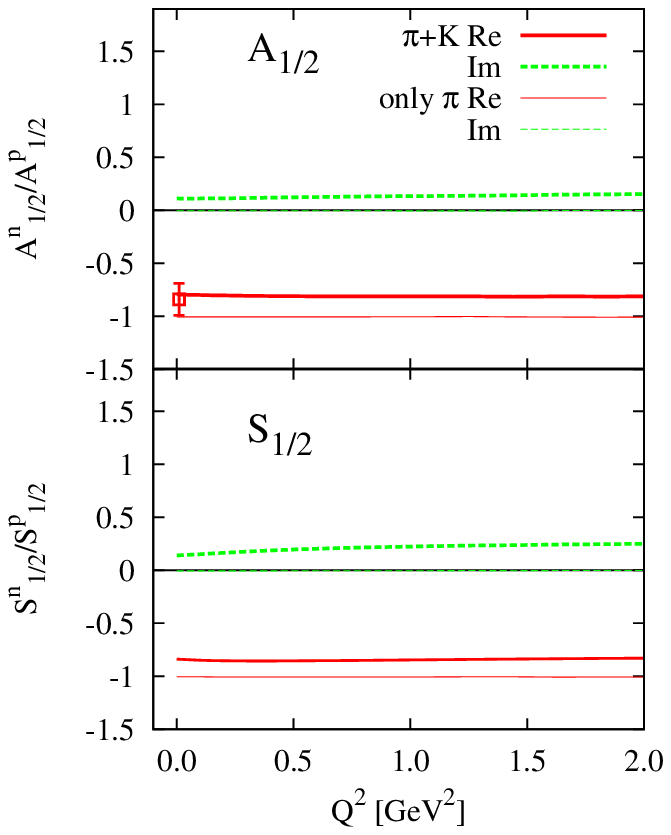}
\caption{Left: Modulus of the $S_{1/2}$ helicity amplitude for the proton resonance as a function 
of $Q^{2}$ with $W=1535 \MeV$. The solid (dotted) line shows the $S_{1/2}^{p}$ amplitude with 
(without) the form factor of the meson inside the loops. 
The sign of this amplitude  relative to $A_{1/2}^p$ is negative, both in experiment and theory.
Right: The $n/p$ ratios of the helicity amplitudes with $W=1535$ MeV. 
   The upper and lower panels show the $np$ ratios of the $A_{1/2}$ and $S_{1/2}$ amplitudes, respectively.
   The open square shows the value given in Ref. \cite{Mukhopadhyay:1995cr}.
\label{fig:S1/2proton}}
\end{figure}

 The results obtained can be seen in figs.  \ref{fig:A1/2proton},
   \ref{fig:S1/2proton}. We can see there
  the $A_{1/2}$ and $S_ {1/2}$ form factors for the proton and neutron 
  as a function of $Q^2$ compared with experiment.  We have done both a
  relativistic and a non relativistic calculation. In both cases the results lye
  a bit below the data. The $Q^2$ dependence, when the results are 
  scaled to the experimental values, is
  fair, particularly for the relativistic case, see fig. \ref{fig:A1/2proton}
  (right). We should note,
  however, that the experimental values are obtained from measured cross
  sections making particular assumptions of the $N^*(1535)$ width and branching
  ratio to $\eta p$ and these uncertainties are not reflected in the data shown
  in the figures. It is instructive to mention that in the latest version of the
  MAID2007 global analysis of data \cite{maid} the value of the $A_{1/2}$ amplitude for real
  photons has the value $66~10^{-3}~GeV^{-1/2}$, which is in 
  agreement with our results.
  
  In any case it is also interesting to observe that the ratio of neutron to
  proton amplitudes, which is free of the global normalization, appears in our
  approach in good agreement with the data.

\section{Photoproduction of $\omega$ and $\omega$ in the nuclear medium}

The interaction of vector mesons with nuclei has captured for long the 
attention of the hadron community. 
Along these lines, an approach has been followed by the CBELSA/ TAPS
collaboration by looking at the $\gamma \pi ^0$ coming from the $\omega$
decay,  where a recent work \cite{trnka} claims evidence for a decrease of the
$\omega$ mass in the medium of the order of 100 MeV from the study of the
modification of the mass spectra in $\omega$ photoproduction. 
Here we present the reanalysis of the data of \cite{trnka} done in
\cite{muratmass}, where one  
 concludes that the distribution is compatible with
an enlarged $\omega$ width of about 90 MeV at nuclear matter density and no
shift in the mass and at the same time we show the insensitivity of the results
to a mass shift.
We also show results for the $(\gamma ,p)$ reaction searching for possible
$\omega$ bound states in the nucleus concluding that even in the case of a
sufficiently attractive potential and small width no peaks can be seen with the
present experimental resolution of about $50 MeV$  at ELSA. We also discuss the
origin of a two peak structure of the $(\gamma ,p)$ cross section which should
not me misidentified with evidence for an $\omega$ bound state in the nucleus.


\section{Preliminaries}

We consider the  photonuclear reaction 
$A(\gamma,\omega\to \pi^0\gamma)X$ in two
steps - 
production of the $\omega$-mesons and 
propagation of the final states.
In the laboratory, where 
the nucleus with the mass number $A$ is at rest, 
the nuclear total  cross section  of the inclusive reaction $A(\gamma,\omega)X$,
including the effects of Fermi motion and Pauli blocking, plus effects of final
state interaction of the particles produced, can be calculated as
shown in \cite{muratmass}.

The $\omega$-mesons are produced according to their
spectral function $S_{\omega}$ at a local density $\rho(r)$
\begin{eqnarray}
\label{SF}
S_{\omega}({m}_{\omega},\widetilde{m}_{\omega},\rho) = 
\hspace{5cm}\nonumber \\
- \frac{1}{\pi} 
\frac{\mbox{Im}\Pi_{\omega}(\rho)}
{\Big(\widetilde{m}_{\omega}^2-{m}_{\omega}^2
-\mbox{Re}\Pi_{\omega}(\rho)\Big)^2 + 
\Big(\mbox{Im}\Pi_{\omega}(\rho)\Big)^2},
\end{eqnarray}
where $\Pi_{\omega}$ is the in-medium selfenergy of the 
$\omega$. The width of the 
$\omega$ in the nuclear medium is 
related to the selfenergy by 
$\Gamma_{\omega}(\rho,\widetilde{m}_{\omega}) 
= - \mbox{Im}\Pi_{\omega}(\rho,\widetilde{m}_{\omega})/E_{\omega}$.
It includes
the free width $\Gamma_{free} = 8.49$~MeV and an in-medium 
part $\Gamma_{coll}(\rho)$ which accounts for the 
collisional broadening of the $\omega$ due to the quasielastic and
absorption channels. In Eq.~(\ref{SF})
$\mbox{Re}\Pi_{\omega}=2 E_{\omega} \mbox{Re}V_{opt}(\rho)$, where
$V_{opt}(\rho)$ is the $\omega$ nucleus optical potential accounts 
for a possible
shift of the $\omega$ mass in the medium and we shall make some considerations
about it latter on.

We also consider the situation when 
the energy of the incident photon beam  is not fixed but 
constrained in some energy interval
$E_{\gamma}^{\min} < E_{\gamma} < E_{\gamma}^{\max}$, and also take into account
the photon flux produced at the ELSA facility.

\section{The Monte Carlo simulation procedure}

The computer MC simulation proceeds in close analogy to the actual
experiment.
At first,
the multiple 
integral involved in the evaluation of the cross section is carried out using 
the MC integration 
method. This procedure provides
 a random 
point $\vec{r}$ inside the nucleus where the photon collides 
with the nucleon, 
also randomly generated
from the Fermi sea with $|\vec{p}_N| \le k_F(\vec{r})$.
For the sample event in the MC integral
the mass $\widetilde{m}_{\omega}$ of the $\omega$
respects 
the spectral function $S_{\omega}$ at local density $\rho(r)$, 
see Eq.~(\ref{SF}).
Inside the nucleus the $\omega$-mesons moving with the three
momentum $\vec{p}_{\omega}^{\,lab}$ necessarily interact with
the nucleons in their way out of the nucleus. In the MC simulation
the $\omega$-mesons are allowed to propagate  
a distance 
$\delta \vec{L} = \frac{\vec{p}_{\omega}^{\,lab}}{|\vec{p}_{\omega}^{\,lab}|}
\delta L$ and at each step, $\delta L \simeq 0.1$~fm, the 
reaction probabilities for different 
channels like
the decay of the $\omega$
into $\pi^0 \gamma$ and $\pi\pi\pi$ final states, 
quasielastic scattering and in-medium absorption
are properly calculated. Details of
the simulation can be seen in \cite{muratmass}.

 We use the following parameterization for the width,
$\Gamma_{abs} = \Gamma_0 \frac{\rho(r)}{\rho_0}$,
where $\rho_0=0.16$~fm$^{-3}$ is the normal nuclear matter density.

The propagation of pions in nuclei is done using a MonteCarlo simulation
procedure . In their way out
of the nucleus pions can experience the quasielastic scattering or 
can be absorbed. 
The intrinsic probabilities for these reactions
as a function of the nuclear matter density
are calculated using the phenomenological model of 
Refs~\cite{simulation},
which also includes 
higher order quasielastic cuts and 
the two-body and three-body
absorption mechanisms. Details for the present case are described in
\cite{muratmass}.

\section{In-medium $\omega$-meson width and nuclear transparency}

In this section we discuss an extraction of the 
in-medium inelastic width of the
$\omega$ in the photonuclear experiments.
As a measure for the $\omega$-meson width in nuclei we employ the so-called
nuclear transparency ratio
\begin{equation}
\tilde{T}_{A} = \frac{\sigma_{\gamma A \to \omega X}}{A \sigma_{\gamma N \to \omega X}}
\end{equation}
i.e. the ratio of the nuclear $\omega$-photoproduction cross section
divided by $A$ times the same quantity on a free nucleon. 
$\tilde{T}_A$ describes
the loss of flux of $\omega$-mesons in the nuclei and is related to the
absorptive part of the $\omega$-nucleus optical potential and thus to the
$\omega$ width in the nuclear medium.

We have done the MC calculations for the
sample nuclear targets:   ${}^{12}_6$C, ${}^{16}_{8}$O,
${}^{24}_{12}$Mg,  ${}^{27}_{13}$Al, ${}^{28}_{14}$Si,
${}^{31}_{15}$P,   ${}^{32}_{16}$S,  ${}^{40}_{20}$Ca,
${}^{56}_{26}$Fe, ${}^{64}_{29}$Cu,  ${}^{89}_{39}$Y, 
${}^{110}_{48}$Cd,  ${}^{152}_{62}$Sm,  ${}^{208}_{82}$Pb,
${}^{238}_{92}$U. 
 In the following we evaluate the ratio between the nuclear 
cross sections in
heavy nuclei and a light one, for instance  $^{12}$C, since in
this way, many other nuclear effects not related to the
absorption of the $\omega$ cancel in the
ratio, $T_A$.

\begin{figure*}[t]
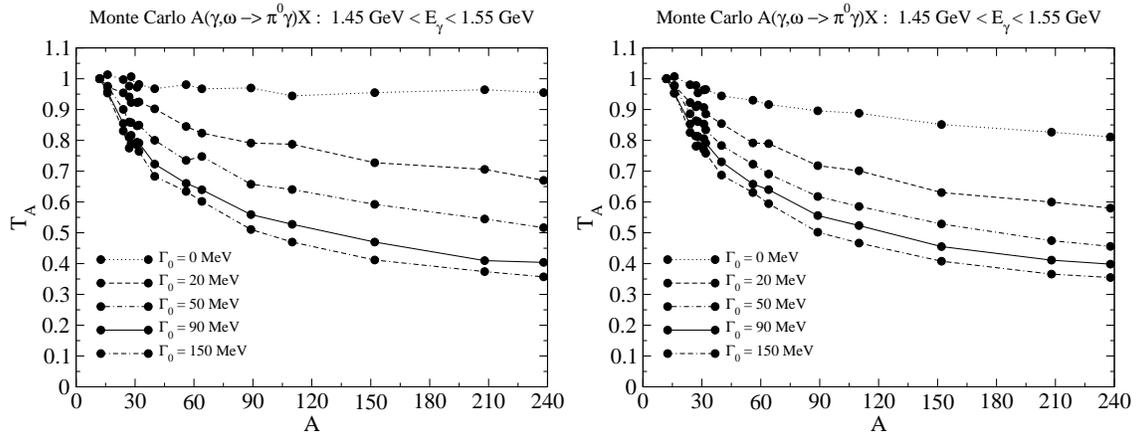

\begin{center}
\includegraphics[clip=true,width=0.40\columnwidth,angle=0.]
{TransparencyMC.eps}
\includegraphics[clip=true,width=0.40\columnwidth,angle=0.]
{TransparencyMCpionCut.eps}
\caption{\label{FigTrMC}  \footnotesize
The result of the Monte Carlo method for the $A$-dependence
of the nuclear transparency ratio $T_A$ without (left panel) and with
(right panel) FSI of outgoing pions. A lower cut $T_{\pi} > 150$~MeV 
on the kinetic energy of the outgoing pions has been used to suppress 
the contribution of the distorted events due to FSI.
The incident photon beam was constrained in the range
$1.45~\mbox{GeV}< E_{\gamma} < 1.55~\mbox{GeV}$.
 The carbon $^{12}$C
was used as the reference target in the ratio of the nuclear cross sections.
With
$\Gamma_{abs} = \Gamma_0 \frac{\rho(r)}{\rho_0}$, where $\rho_0$ is 
the normal nuclear matter density, the dotted, dashed, dash-dotted,
solid and dash-dash-dotted  
curves 
correspond to $\Gamma_0 = 0$~MeV, $\Gamma_0=20$~MeV, $\Gamma_0=50$~MeV, 
$\Gamma_0=90$~MeV and $\Gamma_0=150$~MeV, respectively. 
}
\end{center}
\end{figure*}



The results of the MC calculation
for the $A$-dependence
of the nuclear transparency ratio $T_A$ are presented in Fig.~\ref{FigTrMC}.
The incident photon beam was constrained in the range
$1.45~\mbox{GeV}< E_{\gamma} < 1.55~\mbox{GeV}$ - a region which is considered
in the analysis of the CBELSA/TAPS
experiment~\cite{trnkathesis,Kotulla:2006wz}. 
In Fig.~\ref{FigTrMC} (left panel) we show the results for the
transparency ratio when the collisional broadening and FSI of the $\omega$
are taken into account but without FSI
of the pions from $\omega \to \pi^0 \gamma$ decays inside the nucleus. The right
panel corresponds to considering in addition the FSI of the pions.

By using these results and taking into account
the preliminary results of CBELSA/TAPS experiment~\cite{Kotulla:2006wz} 
we get an estimate 
for the $\omega$ width $\Gamma_{abs} \simeq 90 \times
\frac{\rho(r)}{\rho_0}~\mbox{MeV}$.
This estimate must be understood as an average over the 
$\omega$ three momentum.

\section{In-medium $\omega$-meson mass and CBELSA/TAPS experiment}

The first thing one should note is that the $\omega$ line shape reconstructed
from $\pi^0\gamma$ events strongly depends
on the background shape subtracted from the bare $\pi^0\gamma$ signal.
In Ref.~\cite{trnka} the shape of the background was chosen such that it
accounted for all the experimental strength at large invariant masses.
This choice was done both for the elementary $\gamma p \to  \pi^0 \gamma p$
reaction as well as for nuclei. As we shall show, this choice of background in
nuclei implies a change of the shape from the elementary reaction to that in
the nucleus for which no justification was given. We shall also show
that when the
same shape for the background as for the elementary reaction is chosen, 
the experiment in nuclei shows strength at invariant masses higher than 
$m_{\omega}$ where the choice of~\cite{trnka} necessarily produced no
strength. We will also see that the experimental data can be naturally
interpreted in terms of the large in-medium $\omega$ width discussed above
without
the need to invoke a shift in the $\omega$ mass in the medium. 

\begin{figure}[t]
\begin{center}
\includegraphics[clip=true,width=0.80\columnwidth,angle=0.]
{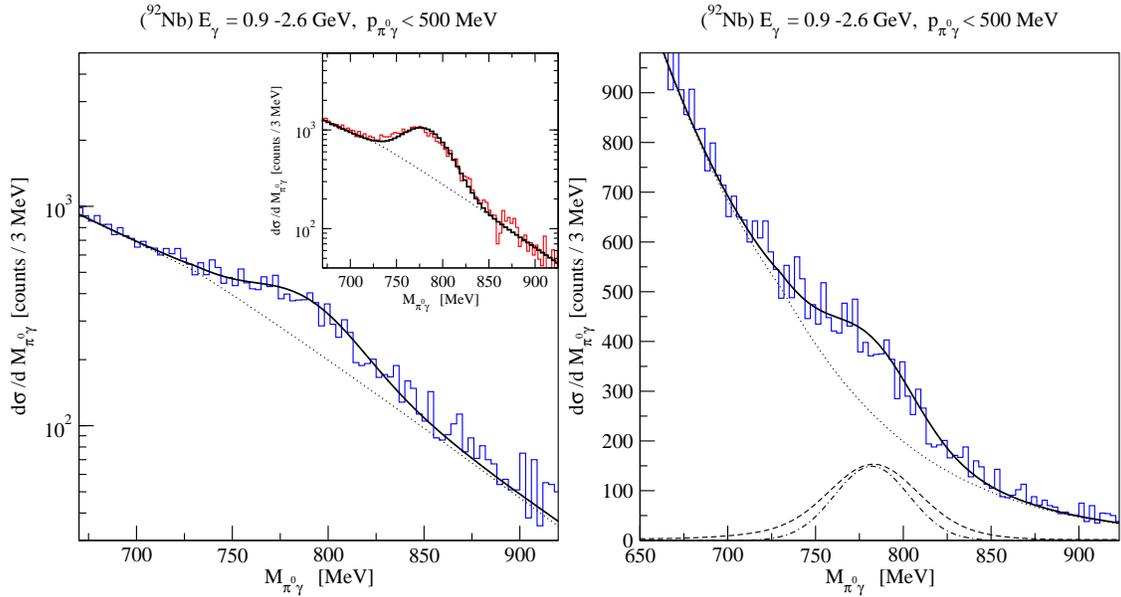}
\caption{\label{PRL}  \footnotesize
Left panel: Invariant mass spectra reconstructed from the $\pi^0\gamma$
events in the $(\gamma,\pi^0\gamma)$ reaction from Nb target (solid curve). 
The experimental data are from Ref.~\cite{trnka}. 
The incident photon beam has been constrained
in the range $0.9$~GeV $< E_{\gamma}^{in} <$ $2.6$~GeV.
Dotted curve is an uncorrelated $\pi^0\gamma$
background (see the text). 
Right panel: Same but in linear scale. 
The dashed and dash-dash-dotted curves correspond to the
$\omega \to \pi^0\gamma$ events with and without the kinematic cut
$|\vec{p}_{\pi^0\gamma}| < 500$ MeV,
respectively. The normalization without cut is arbitrary.
The solid line corresponds to the sum of the
background and the dashed line.
Inset (left panel): The $\pi^0\gamma$ invariant mass spectra in the elementary
$p(\gamma,\pi^0\gamma)p$ reaction. 
Same background line shape (dotted curve) as for the Nb target has been used.
The solid line is the sum of the
background and $\omega \to \pi^0\gamma$ events.}
\end{center}
\end{figure}

In Fig.~\ref{PRL}  
we show the experimental data (solid histogram) for the 
$\pi^0\gamma$ invariant mass spectra in the reaction
$(\gamma,\pi^0\gamma)$~\cite{trnka} from $^{92}_{41}$Nb target. 
The inset (left panel) corresponds to the $\pi^0\gamma$ spectra from the
hydrogen target. 
In our MC calculations the incident photon beam has been constrained
in the range $0.9$~GeV $< E_{\gamma}^{in} <$ $2.6$~GeV. 
The higher momentum cut 
$|\vec{p}_{\pi^0\gamma}| = |\vec{p}_{\pi^0}+\vec{p}_{\gamma}|< 500$~MeV
on a three momentum of the $\pi^0 \gamma$ pair
was imposed as in the actual experiment. First, we use the hydrogen target,
see inset in Fig.~\ref{PRL} (left panel), 
to fix the contribution of the uncorrelated $\pi^0\gamma$
background  (dotted curve) which together with
the $\pi^0\gamma$ signal from $\omega \to \pi^0\gamma$ decay,
folded with the Gaussian experimental resolution of 55 MeV as 
in Ref.~\cite{trnka}, gives a fair
reproduction of the experimental spectra. Then we assume the same 
shape of the $\pi^0\gamma$ 
background in the photonuclear reaction.
The weak effect of the FSI of the pions found in the calculation, with the cuts
imposed in the experiment, strongly supports this assumption.

In the following we use the $\omega$ inelastic 
width of $\Gamma_0 = 90$~MeV
at $\rho_0$.
The exclusive
$\omega \to \pi^0\gamma$ MC spectra is shown by the dashed curve 
(right panel). 
The solid curve is the reconstructed
$\pi^0\gamma$ signal after applying the cut on $\pi^0\gamma$ momenta and
adding
the background fixed when using the hydrogen target (dotted curve). 
Note that the shape of the exclusive $\pi^0 \gamma$ signal without applying 
a cut on $\pi^0\gamma$ momenta
(dash-dotted curve) is dominated by the experimental resolution 
and no broadening of the $\omega$ is observed. This is in agreement
with data of Ref.~\cite{trnka}.
But applying the cut one increases
the fraction of in-medium decays coming from the interior of the nucleus 
where the spectral function is rather broad
and as a result the broadening of the $\pi^0\gamma$ signal with respect
to the signal (without cut) can be well seen.
The resulting MC spectra (solid curve) 
shows the accumulation of the $\pi^0\gamma$ events from the left and right 
sides
of the mass spectra, and it is consistent both with 
our choice of the uncorrelated $\pi^0\gamma$
background and experimental data.

We have also done the exercise of seeing the sensitivity of the results to 
changes in the mass. As shown in \cite{muratmass}, a band corresponding to
having the $\omega$ mass in between $m_{\omega}\pm 40 \rho/\rho_0$~MeV is far
narrower than the statistical fluctuations.
In other words, this 
experiment is too insensitive to changes in the mass to be used for a precise
determination of the shift of the $\omega$-mass in the nuclear medium.8
We should also note that the peak position barely moves since it is
dominated
by the decay of the $\omega$ outside the nucleus.

\section{Production of bound $\omega$ states in the ($\gamma$,p)  
reaction}

Here we evaluate the formation rate of  $\omega$ bound states in the 
nucleus by means of the ($\gamma$,p) reaction.  We use the  
Green function method \cite{NPA435etc} to calculate the cross  
sections for $\omega$-mesic states formation as described in
Refs.~\cite{NPA761} in   
detail.
The theoretical model used here is exactly same as that used in these
references.

The $\omega$-nucleus optical potential is written here as 
$V(r) = (V_0 + iW_0  ) \frac{\rho(r)}{\rho_0}$, 
where  $\rho(r)$ is the nuclear experimental density for which we take 
 the two parameter Fermi  distribution.
We consider three cases of the potential strength as: $(V_0,W_0) = -(0,50)$,
$-(100,50)MeV$ and $-(156,29) MeV$.
The last of the potentials is
obtained by the linear density approximation with the scattering length
$a=1.6 + 0.3 i$ fm \cite{klingl2}.
This potential is strongly attractive with weak
absorption and hence should be the ideal case for the formation
of $\omega$ mesic nuclei.
No $\omega$ bound states are expected for the first  potential 
which has only an absorptive part.
The second potential  has a strong attraction 
with the large absorptive part as indicated in
Ref.~\cite{trnkathesis}.
For the first two potentials we find no visible peaks in the
spectrum since the width is so large.
For the third potential we observe peaks but they are washed out when folded 
with the experimental resolution of about $50 MeV$ of ELSA.

\section{Monte Carlo simulation of the reaction of the $(\gamma,p)$ reaction}

We next apply the MonteCarlo simulation explained above to describe the 
$(\gamma,p)$ reaction studied at ELSA.
Because our MC calculations represent complete event simulations
it is possible to take into account the actual experimental 
acceptance of ELSA~\cite{trnkathesis} (see details in \cite{hidekoatoms}).

We start our MC analysis with the cross section of 
the elementary reaction $\gamma p \to \omega p \to \pi^0\gamma p$. With this we
determine the cross section for $\omega$ formation and follow the fate of the
protons at the same time.

\begin{figure*}[t]
\begin{center}
\includegraphics[clip=true,width=0.50\columnwidth,angle=0.]
{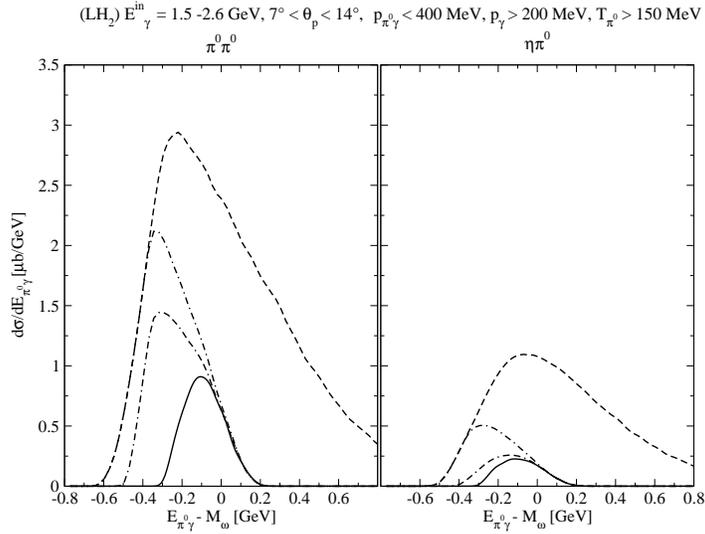}
\caption{\label{FigNewBGR} \small
The differential cross section $d\sigma/dE_{\pi^0 \gamma}$ 
of the reactions $\gamma p \to \pi^0 \pi^0  p$ (left panel)
and $\gamma p \to \pi^0 \eta  p$ (right panel) followed by the decay
$\pi^0(\eta) \to \gamma \gamma$
as a function of the
$E_{\pi^0\gamma}-m_{\omega}$ where $E_{\pi^0\gamma}=E_{\pi^0}+E_{\gamma}$.
 The following cuts were imposed: 
$E_{\gamma}^{in}= 1.5 \div 2.6$~GeV and $7^{\circ} < \theta_p < 14^{\circ}$
(dashed curves);
$E_{\gamma}^{in}= 1.5 \div 2.6$~GeV, $7^{\circ} < \theta_p < 14^{\circ}$
and $|\vec{p}_{\pi^0}+\vec{p}_{\gamma}| < 400$~MeV (dash-dotted curves);
plus  the cut $T_{\pi^0} > 150$~MeV (dash-dash-dotted curves)
 and plus the cut $|\vec{p}_{\gamma}| > 200$~MeV (solid curves).
}
\end{center}
\end{figure*}

There are  also sources of background like from $\gamma p \to \pi^0
\pi^0 p$, or $\gamma p \to \pi^0 \eta p$, where one of the two photons from
the decay of the $\pi^0$ or the $\eta$ is not measured. 
 We show in Fig.~\ref{FigNewBGR} 8
the cross section 
$d \sigma /d E_{\pi^0 \gamma}$ coming from the 
$\gamma p \to \pi^0\pi^0 p$ reaction 
followed by the decay $\pi^0 \to \gamma \gamma$ of either of the 
$\pi^0$ (left panel)
and from the $\gamma p \to \pi^0\eta p$ reaction followed by the decay 
$\eta \to \gamma \gamma$ (right panel). 
  As one can see, the
contribution from the $\pi^0 \pi^0$ 
photoproduction to the background is the dominant
one among the two.  The important thing, thus, is that these two 
sources of background, with the cuts imposed,  produce a background 
peaked at -100 MeV. For the exclusive $\pi^0 \gamma$ 
events coming from $\gamma p \to \omega p \to \pi^0\gamma p$ 
an experimental resolution of $50$~MeV was imposed, 
see Ref.~\cite{trnka}. 
We obtain a factor of two bigger strength
at the $\omega$ peak than at the peak from the $\gamma p \to \pi^0 \pi^0 p$
background. Experimentally, this seems to be also
the case from the preliminary data of CBELSA/TAPS,

\begin{figure*}[t]
\begin{center}
\includegraphics[clip=true,width=0.45\columnwidth,angle=0.]
{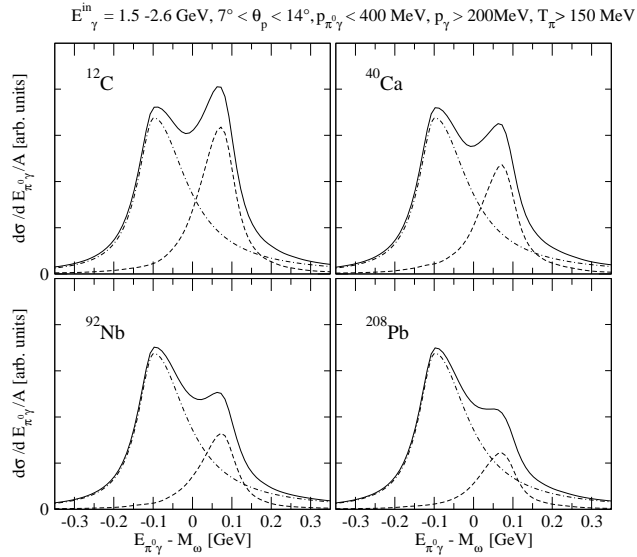}
\caption{\label{FigEMnuclear} \small
The differential cross section $d\sigma/dE_{\pi^0\gamma}$ 
of the reaction $A(\gamma,\pi^0\gamma)X$
as a function
of $E_{\pi^0\gamma} - m_{\omega}$  from $^{12}$C, $^{40}$Ca, $^{92}$Nb and
$^{208}$Pb nuclear targets. The reconstructed exclusive events from the 
$\omega \to \pi^0 \gamma$ decay are
shown by the dashed curves.  The $\pi^0 \gamma$ background is shown by
the dash-dotted curves. 
The sum of the two contributions is given by the solid curves.
The following cuts were imposed: 
$E_{\gamma}^{in}= 1.5 \div 2.6$~GeV, $7^{\circ} < \theta_p < 14^{\circ}$,
$|\vec{p}_{\pi^0}+\vec{p}_{\gamma}| < 400$~MeV, $|\vec{p}_{\gamma}| > 200$~MeV 
and $T_{\pi} > 150$~MeV. The exclusive $\omega \to \pi^0 \gamma$ signal 
has been folded with the 50 MeV experimental resolution.
All spectra are normalized to the corresponding 
nuclear mass numbers $A$.} 
\end{center}
\end{figure*}


In the following we assume that the inclusive
$\pi^0 \gamma$ background scales with respect to 
the target nucleus 
mass number $A$ like $\sigma_A \simeq A \, \sigma_{elem}$.
But this is not
the case for the exclusive $\pi^0 \gamma$ events coming from the decay of the
$\omega \to \pi^0 \gamma$, since 
the rather strong absorption
of the $\omega$ inside the nucleus changes the scaling relation
and $\sigma_A(\omega \to \pi^0 \gamma) \simeq A^{\alpha} \, 
\sigma_{elem}(\omega \to \pi^0 \gamma)$,
 where the attenuation parameter $\alpha < 1$.

In Fig.~\ref{FigEMnuclear} 
we show the result of the MC simulation for the 
$E_{\pi^0 \gamma} - m_{\omega}$ 
spectra reconstructed from the $\pi^0$ and $\gamma$ events. The calculations
are performed 
for the sample nuclear targets $^{12}\mbox{C}$, $^{40}\mbox{Ca}$, 
$^{92}\mbox{Nb}$ and $^{208}\mbox{Pb}$. The kinematic and acceptance
cuts discussed before  have been already imposed. 
The MC distributions are
normalized to the nuclear mass number $A$. The solid curves correspond
to the sum of the inclusive $\pi^0 \gamma$ background (dash-dotted curve), 
 and the exclusive $\pi^0 \gamma$ 
events coming from the direct
decay of the $\omega\to \pi^0 \gamma$. The contributions of the exclusive
$\omega \to \pi^0 \gamma$
events are shown by the dashed curves. We note a
very strong attenuation of the
$\omega\to \pi^0 \gamma$ signal with respect to the background contribution
 with increasing  nuclear mass number $A$.
 This is
primary due to the stronger absorption of the $\omega$-mesons 
with increasing nuclear matter density.
  The former exercise indicates that given the particular combination of 
$\pi^0 \gamma$ from an uncorrelated background and from  $\omega$ decay, and
the different behaviour of these two sources in the $\pi^0 \gamma$ production in
nuclei, a double hump structure is unavoidable in nuclei with this set up, and 
one should avoid any temptation
to associate the lower energy peak to a possible bound state in the nucleus. 

\section{\label{Summary} Conclusions}

In this lecture we have addressed several recent topics on photoproduction of
mesons on nucleons and nuclei. One of the topics was the combined study of the
 $\gamma p \to \pi^0 \eta p$ and $\gamma p \to \pi^0 K^0 \Sigma^+$ reactions,
 which, together with related pion induced reactions, showed a dominance of the
 $\Delta(1700)$ excitation. This resonance is one of the dynamically generated
 resonances in the chiral unitary approach and this theory provides couplings
 unaccessible through known decay widths which allow a global understanding of
 these reactions. 
 
  The study of the $e^+ e^- \to \phi f_0(980)$ reaction, using again techniques
  of chiral unitary theory, allowed us to calculate the bulk of the cross section
  for this reaction, which was well reproduced, and comparison with data showed
  the need to introduce a new resonance, the X(2175), which has been
  claimed from the analysis of a recent BABAR experiment.
  
  Similarly, using information obtained in the chiral approach to the
  $N^*(1535)$, we showed results obtained for the electromagnetic form factors
  of this resonance, which provide a fair reproduction of the $Q^2$ dependence,
  and a normalization for real photons in agreement with the MAID2007 analysis,
  though a little lower than experiments without counting the systematic
  experimental uncertainties in the normalization.
  
 Concerning photoproduction$\omega$, the studies done in \cite{muratmass} and \cite{hidekoatoms} show that: 1) The
 ELSA results on inclusive $\omega$ production in nuclei can be interpreted in
 terms of a large $\omega$ width in the medium without the need of a mass shift.
  2) The results are very insensitive to a mass shift in  matter. 3) With the
  large medium $\omega$ width derived from the ELSA data no visible peaks for
  $\omega$ bound states are seen, even with hypothetical large $\omega $
  binding. 4) Even in the hypothetical case of small widths, the possible
  $\omega$ bound states would not be resolved with the present ELSA resolution.
  5) When looking at the $(\gamma,p)$ reaction with the present ELSA
  experimental set up, a double hump structure appears in the calculation from
  the interplay of the $\omega$ signal and the background. The peak at lower
  energies is related to the background, with the cuts imposed, and should not
  me misidentified with a possible $\omega$ bound state in the nucleus. \\
 
{\bf Note added in Proofs:} \\

 After thorough discussions with V. Metag and M. Kotulla it became clear that
 the issue of using the mixed event technique (MET) to get the background in
 $\omega$ production in proton and nuclear targets is very delicate. Questions
 like normalizations, effects of cuts, comparison of the MET results on proton
 and nuclear targets, and particularly, the ability of the MET to generate the
 background in a very correlated system like in the present case, are far from
 settled. In this respect, the results on the MET reported in the contribution
 of V. Metag in this School [arxiv 0711.4709 , nucl-ex] should be  taken as
 very preliminary, and as such, one should abstain from drawing conclusions
 from them.

    Another topic that requires clarification is the comment in the same
contribution which states " since the experimental data clearly show that the
background distributions on the proton and in nuclear targets are  different,
the assumption made in section 11 of the present paper (and references  quoted
there) of a background in the nucleus proportional to the one of the proton can
only lead to wrong conclusions". This statement is incorrect for several 
reasons: 1) The experimental data quoted above are for total cross sections and
do  not tell us which is the background. 2) In the work reported in section 11
of  the present paper one does not claim that the background in the nucleus is 
proportional to the one of the proton. It shows it as a possible scenario and
discusses what would happen in such a case. 3) Although we do not necessarily
advocate this latter scenario, it is impossible to exclude it from the present
data alone, since in ref. [56] it was proved that it is indeed one of the
possibilities for  a certain $\omega$ signal in the nucleus.

\section*{Acknowledgments}  
This work is partly supported by DGICYT contract
number FIS2006-003438, the Generalitat Valenciana, the projects FPA2004-05616
(DGICYT) and SA016A07 (Junta de Castilla y Leon). The work of M.N. and C.A.V.A.
 was supported by CONACyT- Mexico under project 50471-F. 
 This research is  part of the EU Integrated
Infrastructure Initiative  Hadron Physics Project under  contract number
RII3-CT-2004-506078.


\begin{thebibliography}{99}
\itemsep -2pt 

 \bibitem{xpt} J.~Gasser and H.~Leutwyler,
Nucl.\ Phys.\  {\bf B250} (1985) 465, 517, 539.


\bibitem{ulf}U. G. Meissner, Rep. Prog. Phys. {{56}} (1993)
903; V. Bernard, N. Kaiser and U. G. Meissner, Int. J. Mod. Phys. {{E4}} (1995)
193.

\bibitem{ecker}G. Ecker, Prog. Part. Nucl. Phys. {{35}} (1995) 1.

\bibitem{Veit:1984jr}
  E.~A.~Veit, B.~K.~Jennings, A.~W.~Thomas and R.~C.~Barrett,
  Phys.\ Rev.\  D {\bf 31} (1985) 1033.

\bibitem{Kaiser:1995eg}
  N.~Kaiser, P.~B.~Siegel and W.~Weise,
  Nucl.\ Phys.\ A {\bf 594} (1995) 325
  

\bibitem{kaon} E. Oset and A. Ramos,  \NPA{635} (1998) 99.


\bibitem{Oller:2000fj}
J.~A.~Oller and U.~G.~Meissner,
Phys.\ Lett.\ B {\bf 500} (2001) 263

\bibitem{nsd} J. A. Oller and E. Oset, \PRD{60} (1999) 074023.






\bibitem{Kaiser:1995cy}
N.~Kaiser, P.~B.~Siegel and W.~Weise,
Phys.\ Lett.\ B {\bf 362} (1995) 23





\bibitem{Nieves:2001wt}
J.~Nieves and E.~Ruiz Arriola,
Phys.\ Rev.\ D {\bf 64} (2001) 116008



\bibitem{Inoue:2001ip}
T.~Inoue, E.~Oset and M.~J.~Vicente Vacas,
Phys.\ Rev.\ C {\bf 65} (2002) 035204


\bibitem{review} J. A. Oller, E. Oset and A. Ramos, Prog. Part. Nucl. Phys. 45
(2000) 157.

\bibitem{GarciaRecio:2003ks}
  C.~Garcia-Recio, M.~F.~M.~Lutz and J.~Nieves,
  Phys.\ Lett.\  B {\bf 582} (2004) 49
 
  
\bibitem{GarciaRecio:2005hy}
  C.~Garcia-Recio, J.~Nieves and L.~L.~Salcedo,
  Phys.\ Rev.\  D {\bf 74} (2006) 034025
  
  
\bibitem{Hyodo:2002pk}
  T.~Hyodo, S.~I.~Nam, D.~Jido and A.~Hosaka,
  Phys.\ Rev.\  C {\bf 68} (2003) 018201
  
  
\bibitem{Hyodo:2006yk}
  T.~Hyodo, D.~Jido and A.~Hosaka,
  Phys.\ Rev.\ Lett.\  {\bf 97} (2006) 192002
 

\bibitem{bennhold} E.~Oset, A.~Ramos and C.~Bennhold,
Phys.\ Lett.\ B {\bf 527} (2002) 99
[Erratum-ibid.\ B {\bf 530} (2002) 260].











\bibitem{Jido:2003cb}
D. Jido, J.A. Oller, E. Oset, A. Ramos, U.G. Meissner,
Nucl.\ Phys.\ A {\bf 725} (2003) 181.


 
\bibitem{pramana}
  E.~Oset, D.~Cabrera, V.~K.~Magas, L.~Roca, S.~Sarkar, M.~J.~Vicente Vacas and A.~Ramos,
  Pramana {\bf 66} (2006) 731
  
  
 \bibitem{lutz}
E.~E.~Kolomeitsev and M.~F.~M.~Lutz,
Phys.\ Lett.\ B {\bf 585} (2004) 243

\bibitem{sarkar}
  S.~Sarkar, E.~Oset and M.~J.~Vicente Vacas,
  Nucl.\ Phys.\  A {\bf 750} (2005) 294
  [Erratum-ibid.\  A {\bf 780} (2006) 78]
  
  
\bibitem{Jenkins:1991es}
E.~Jenkins and A.~V.~Manohar,
Phys.\ Lett.\ B {\bf 259} (1991) 353.



\bibitem{bookericson} Pions and Nuclei, T. E. O. Ericson and W. Weise, Oxford
Science Publications, 1988.

\bibitem{holstein}
T.~R.~Hemmert, B.~R.~Holstein and J.~Kambor,
J.\ Phys.\ G {\bf 24} (1998) 1831



\bibitem{savage} M. Savage, http://www.phys.washington.edu/$\sim$savage/


\bibitem{lutz3}
M.~F.~M.~Lutz and E.~E.~Kolomeitsev,
Nucl.\ Phys.\ A {\bf 700} (2002) 193



\bibitem{Roca:2006sz}
  L.~Roca, S.~Sarkar, V.~K.~Magas and E.~Oset,
  Phys.\ Rev.\  C {\bf 73} (2006) 045208
  
  
\bibitem{etapi}
  M.~Doring, E.~Oset and D.~Strottman,
  Phys.\ Rev.\  C {\bf 73} (2006) 045209
  
  
\bibitem{kpi}
  M.~Doring, E.~Oset and D.~Strottman,
  Phys.\ Lett.\  B {\bf 639} (2006) 59
  
  


\bibitem{dahl}
O.~I.~Dahl, L.~M.~Hardy, R.~I.~Hess {\it et.al.}, Phys.\ Rev.\ {\bf 163}, 1337 (1967).
\bibitem{curtis}
L.~J.~Curtis, C.~T.~Coffin, D.~I.~Meyer, and K.~M.~Terwilliger, Phys.\ Rev.\ {\bf 132}, 1771 (1963).
\bibitem{Hanson:1972zz}
  P.~Hanson, G.~E.~Kalmus and J.~Louie,
  Phys.\ Rev.\ D {\bf 4}, 1296 (1971).

\bibitem{Grether:1973sz}
  D.~Grether, G.~Gidal and G.~Borreani,
  Phys.\ Rev.\ D {\bf 7}, 3200 (1973).

  \bibitem{Thomas:1973uh}
  D.~W.~Thomas, A.~Engler, H.~E.~Fisk and R.~W.~Kraemer,
  Nucl.\ Phys.\ B {\bf 56}, 15 (1973).
  

\bibitem{wieland} F. Wieland, private communication.







  

\bibitem{mariana}
M. Nanova at the ''International Workshop On The Physics Of Excited Baryons (NSTAR 05)'',
10-15 Oct 2005, Tallahassee, Florida

\bibitem{nakabayashi}
  T.~Nakabayashi {\it et al.},
  Phys.\ Rev.\  C {\bf 74} (2006) 035202.
  
  
\bibitem{graal} J. Ajaka et al., preprint 2007. Phys. Rev. Lett. in print.
 \bibitem{Erbe:1970cq}
  R.~Erbe {\it et al.}  [Aachen-Berlin-Bonn-Hamburg-Heidelberg-Muenchen
                  Collaboration],
  Phys.\ Rev.\  {\bf 188}, 2060 (1969).
\bibitem{Erbe:2}
  R.~Erbe {\it et al.}  [Aachen-Berlin-Bonn-Hamburg-Heidelberg-Muenchen
                  Collaboration],
  Nuovo\ Cimento\ {\bf 49A}, 504 (1967).
\bibitem{cmc}
Cambridge Bubble Chamber Group, Phys.\ Rev.\ {\bf 156}, 1426 (1966)  

 
 

\bibitem{BBY} 
B.~Aubert \textit{et al.} [BABAR Collaboration], 
Phys.\ Rev.\ Lett.\ \textbf{95}, 142001 (2005) 

\bibitem{BBX} 
B.~Aubert \textit{et al.} [BABAR Collaboration], 
Phys.\ Rev.\ D \textbf{74}, 091103 (2006) ; 
 B.~Aubert {\it et al.}  [BABAR Collaboration],
 Phys.\ Rev.\  D {\bf 76} (2007) 012008
 [arXiv:0704.0630 [hep-ex]].


\bibitem{eugenio} 
E.~Marco, S.~Hirenzaki, E.~Oset and H.~Toki, 
Phys.\ Lett.\ B \textbf{470}, 20 (1999) ; 
J.~E.~Palomar, L.~Roca, E.~Oset and M.~J.~Vicente Vacas, 
Nucl.\ Phys.\ A \textbf{729}, 743 (2003) ; 

  
  
\bibitem{others}
V.~E.~Markushin, 
Eur.\ Phys.\ J.\ A \textbf{8}, 389 (2000); 
  J.~A.~Oller,
  Nucl.\ Phys.\  A {\bf 714}, 161 (2003)
  
  
\bibitem{mauro}
  M.~Napsuciale, E.~Oset, K.~Sasaki and C.~A.~Vaquera-Araujo,
  Phys.\ Rev.\  D {\bf 76} (2007) 074012
 

  
  
\bibitem{Krusche:1995nv}
  B.~Krusche {\it et al.},
  Phys.\ Rev.\ Lett.\  {\bf 74} (1995) 3736.


  
 
\bibitem{Denizli:2007tq}
  H.~Denizli {\it et al.}  [CLAS Collaboration],
  Phys.\ Rev.\  C {\bf 76}, 015204 (2007)
 

\bibitem{Thompson:2000by}
  R.~Thompson {\it et al.}  [CLAS Collaboration],
 Phys.\ Rev.\ Lett.\  {\bf 86}, 1702 (2001)
  


\bibitem{Brasse:1977as}
  F.~W.~Brasse {\it et al.},
  Nucl.\ Phys.\  B {\bf 139}, 37 (1978);
  F.~W.~Brasse {\it et al.},
  Z.\ Phys.\  C {\bf 22}, 33 (1984).

\bibitem{Beck:1974wd}
  U.~Beck {\it et al.},
  Phys.\ Lett.\  B {\bf 51}, 103 (1974).


\bibitem{Breuker:1978qr}
  H.~Breuker {\it et al.},
  Phys.\ Lett.\  B {\bf 74}, 409 (1978).
  


\bibitem{Mukhopadhyay:1995cr}
  N.~C.~Mukhopadhyay, J.~F.~Zhang and M.~Benmerrouche,
  Phys.\ Lett.\  B {\bf 364}, 1 (1995)
  
  
\bibitem{maid}
  D.~Drechsel, S.~S.~Kamalov and L.~Tiator,
  arXiv:0710.0306 [nucl-th].
  
  

  
  \bibitem{trnka}
  D.~Trnka {\it et al.}  [CBELSA/TAPS Collaboration],
  Phys.\ Rev.\ Lett.\  {\bf 94} (2005) 192303
  
 
  
\bibitem{muratmass}
  M.~Kaskulov, E.~Hernandez and E.~Oset,
  Eur.\ Phys.\ J.\  A {\bf 31} (2007) 245
  
  
 

\bibitem{simulation}
  L.~L.~Salcedo, E.~Oset, M.~J.~Vicente-Vacas and C.~Garcia-Recio,
  Nucl.\ Phys.\ A {\bf 484} (1988) 557.

\bibitem{trnkathesis} D. Trnka, PhD Thesis, University of Giessen, 2006.

\bibitem{Kotulla:2006wz}
  M.~Kotulla,
  [arXiv:nucl-ex/0609012].
  
 \bibitem{NPA435etc}
O. Morimatsu, K. Yazaki, Nucl. Phys. A 435 (1985) 727;

 \bibitem{NPA761}
D. Jido, H. Nagahiro, S. Hirenzaki, Phys. Rev. C 66 (2002) 045202;

\bibitem{klingl2}
  F.~Klingl, T.~Waas and W.~Weise,
  Nucl.\ Phys.\ A {\bf 650} (1999) 299
  
  


\bibitem{hidekoatoms}
  M.~Kaskulov, H. Nagahiro, S. Hirenzaki and E.~Oset,
  Phys. Rev. C75 (2007) 064616

  
  
  
  
  
 

  

\end{thebibliography}
\end{document}